\documentclass[%
 reprint,
 amsmath,amssymb,
 aps,
pra,
]{revtex4-1}

\usepackage{braket}
\usepackage{graphicx}% Include figure files
\usepackage{dcolumn}% Align table columns on decimal point
\usepackage{bm}% bold math
\usepackage[normalem]{ulem} % strikethrough
\usepackage{color}

\usepackage{hyperref}% add hypertext capabilities
\hypersetup{
    colorlinks,%
    citecolor=blue,%
    linkcolor=blue,%
    urlcolor=blue
}

\newcommand{\msc}{(MC$_4$S$_4$)$_3$}
\newcommand{\Ni}{(NiC$_4$S$_4)_3$}
\newcommand{\Pt}{(PtC$_4$S$_4)_3$}
\newcommand{\fp}{first-principles}
\begin{document}

%\preprint{APS/123-QED}
\title{Tuning the topological states in {metal-organic} bilayers}

%%\thanks{A footnote to the article title}
%
\author{F. Crasto de Lima, Gerson J. Ferreira, and R. H. Miwa}
\affiliation{Instituto de F\'isica, Universidade Federal de Uberl\^andia, \\
        C.P. 593, 38400-902, Uberl\^andia, MG,  Brazil}%
\date{\today}

\begin{abstract}

We have investigated the energetic stability and the electronic properties of 
metal-organic topological insulators bilayers (BLs), \msc-BL, with M=Ni and Pt, 
using \fp\,  calculations and  tight-binding model. Our findings show that 
\msc-BL is an appealing platform to perform electronic band structure 
engineering, based on the topologically protected chiral edge states.  The 
energetic stability of the BLs is ruled by van der Waals interactions; being  
the AA stacking  the energetically most stable one. The electronic band 
structure is characterized by a combination of bonding and anti-bonding kagome 
band sets (KBSs), revealing that \Ni-BL presents a Z$_2$-metallic phase, whereas 
\Pt-BL  may present both Z$_2$-metallic phase or quantum spin Hall phase. Those 
non-trivial topological states were confirmed by the formation  of chiral edge 
states in \msc-BL nanoribbons. We show that the localization of the edge states 
can be controlled with a normal external electric field, breaking the mirror 
symmetry. Hence, the sign of electric field selects in which layer each set of 
edge states are located. Such a control on the (layer) localization, of the 
topological edge states, bring us an additional and interesting  degree of 
freedom to control the transport properties in layered metal-organic topological 
insulator. 

\end{abstract}

\maketitle

\section{Introduction}

Two dimensional (2D) topological insulators, based on organic hosts, have been 
the  subject of numerous studies addressing not only fundamental issues, but 
also future technological applications. In a seminal work, Wang {\it et al.} 
\cite{wangNatCommun2013} predicted a non-trivial topological phase in an organic 
lattice composed by a monolayer (ML) of three benzene molecules bonded to metal 
atoms, Pb and Bi. Soon after the successful  synthesis of 2D metal-organic 
ML lattices of nickel bis(dithiolene), \Ni\,\cite{kambeJACS2013}, theoretical 
studies based on \fp\, calculations and single orbital tight-binding (TB) model, 
predicted a non-trivial topological phase in \Ni, characterized by the 
topological invariant Z$_2$ [=1 in \Ni], and the formation of spin-polarized 
chiral edge states at the time-reversal-invariant momenta
(TRIM)\,\cite{NANO2842Feng}.

By exploiting the large variety of (possible) combinations of metal-organic 
hosts, other metal-organic frameworks (MOFs), with non-trivial topological 
phase, have been proposed in the past few years. For instance, keeping the 
kagome lattice of \Ni, but substituting Ni with Mn atoms, Zhao {\it et 
al.}\,\cite{zhaoNanoscale2013} verified the quantum anomalous Hall (QAH) state 
in (MnC$_4$S$_4$)$_3$.  Here, the appearance of a ferromagnetic phase, mediated 
by the unpaired Mn-3$d$ electrons, breaks the time-reversal symmetry  of the 
original \Ni\, system. Further QAH state has also been predicted in 2D lattices 
of (i) {\it trans}-Au-THTAP, where the ferromagnetism arise due to a half-filled flat band\,\cite{PRB94Aoki} ; and (ii) triphenil-manganese (MnC$_4$H$_5$)$_3$\,\cite{wangPRL2013}, where ferromagnetically coupled Mn atoms are connected by benzene rings forming a honeycomb lattice. By keeping 
the same honeycomb structure of the benzene host, and substituting Mn with Pb 
atoms (triphenil-manganese$\rightarrow$triphenil-lead), it has been predicted a 
non-magnetic ground state, where the spin-orbit coupling (SOC) promotes  the 
QSH phase in (PbC$_4$H$_5$)$_3$\,\cite{wangNatCommun2013}. Further 
investigations\,\cite{kimPRB2016} pointed out that,   mediated by an external 
electric field,  the (PbC$_4$H$_5$)$_3$ lattice presents an 
energetically stable ferrimagnetic QAH phase. Meanwhile, the recently 
synthesized  Ni$_3$(C$_{18}$H$_{12}$N$_6$)$_2$ MOF\,\cite{sheberlaJACS2014} can 
be considered as the  experimental realization of the so called topological 
Z$_2$-metallic phase\,\cite{panNewJPhys2014} in MOFs.  
It  is characterized by   a kagome lattice, with 
a global energy gap at the edge of the Brillouin zone (K point), whereas the 
energy dispersion of the flat (kagome band) along the $\Gamma$--K direction 
gives rise to a local gap at the $\Gamma$ point\,\cite{PRB90Zhao}.

The design of 2D systems based on the MOFs  is not limited by the 
metal$\leftrightarrow$organic-host combinations. Based on the recent concept of 
van der Waals (vdW) heterostructures\,\cite{geimNature2013},  we may access a 
set of new/interesting electronic properties by stacking 2D MOFs, as we have 
testified in inorganic layered materials\,\cite{radisavljevicNatNanotech2011}. 
 Currently we are facing a suitable  synergy  between the 
experimental  works addressing the successful synthesis of stacked 2D 
MOFs\,\cite{kitagawaAngew2004,colsonNatChem2013,zhangJACS2014, 
sheberlaJACS2014,rodriguezChemComm2016,sakamotoCoord2016}, and  theoretical 
studies aiming the understanding of their physical properties; and propose the 
design of new atomic structures\,\cite{adjizianNatComm2014,zhouRSCAdv2014} 
focusing on a set of desired electronic properties. For instance, the control of 
the topological states in stacked MOFs.

In this paper we investigate  the energetic stability and the electronic 
properties of   \msc\, (M=Ni and Pt)  bilayers, \msc-BLs. The present study was 
carried out through  a combination of \fp\, calculations and  TB model. The 
energetic stability of the \msc-BLs is ruled by vdW interactions; where (i) the 
electronic band structure of the most likely BL configuration (AA stacking) is   
characterized by a combination of bonding and anti-bonding  kagome band sets 
(KBSs).  The non-trivial nature of the  energy gaps, induced by the SOC,  was 
verified through the calculation of the edge states in \msc-BL nanoribbons 
(NRs). (ii) Turning on an external electric field normal to the BL, we find that 
the electronic contributions from each ML are no longer symmetric; giving rise 
to  an interlayer separation between the bonding and anti-bonding KBSs. By  
mapping the localization of the edge states,  we find  that they  follow the 
same spacial separation pattern, showing that the (layer)  localization of the 
topologically protected edge states in \msc-BL NRs can  be tuned by the external 
electric field. Based upon the \fp\, calculations and a phenomenological model, 
we can infer that (i) and (ii), described above, will also take place in other 
vdW metal-organic BLs charaterized by a superposition of kagome bands.

\section{Method}

     The calculations were performed based on the DFT approach, as implemented 
in the VASP code\cite{vasp1}. The exchange correlation term was described using 
the GGA functional proposed by Perdew, Burke and Ernzerhof (PBE)\cite{PBE}. The 
Kohn-Sham orbitals are expanded in a plane wave basis set with an energy cutoff 
of 400 eV. The 2D Brillouin Zone (BZ) is sampled according to the Monkhorst-Pack 
method\cite{PhysRevB.13.5188}, using a gamma-centered 4$\times$4$\times$1 mesh 
for atomic structure relaxation and 6$\times$6$\times$1 mesh to obtain the 
self-consistent total charge density. The electron-ion interactions are taken 
into account using the Projector Augmented Wave (PAW) method 
\cite{PhysRevB.50.17953}. All geometries have been relaxed until atomic forces 
were lower than $0.025$\,eV/{\AA}. The metal-organic framework  monolayer 
system is simulated considering a vacuum region in the direction perpendicular 
to the layers of at least $16$\,{\AA}. For MOF bilayers the van der Waals 
interaction (vdW-DF2\cite{VDW-DF2})  was considered to correctly describe the 
system. In this bilayer system the vacuum region is increased to at least 
$24$\,{\AA} to avoid periodic images interaction.

	The real-space tight-binding (TB) Hamiltonian of kagome-hexagonal 
lattice\,\cite{PRLtang2011, wangPRL2013} in the presence of intrinsic spin-orbit coupling can be written as
	\begin{equation}
	H_{TB} = H_0 + H_{SO} \label{tb}
	\end{equation}
where each term is given by
$$
H_0 = t_1 \sum_{\langle i j\rangle; \alpha} c_{i \alpha}^{\dagger} c_{j \alpha} 
+ t_2 \sum_{\langle \langle i j \rangle \rangle; \alpha} c_{i \alpha}^{\dagger} 
c_{j \alpha} ;
$$

\begin{eqnarray*}
H_{SO} = i\, {\lambda}_{1} \sum_{ \langle i j \rangle} c_{i}^{\dagger} 
\bm{\sigma} \cdot (\bm{d}_{kj} \times \bm{d}_{ik}) c_j + \\ i\, {\lambda}_{2} 
\sum_{\langle \langle i j \rangle \rangle} c_{i}^{\dagger} \bm{\sigma} \cdot 
(\bm{d}_{kj} \times \bm{d}_{ik}) c_j ;
\end{eqnarray*}

		Here, $c_{i \alpha}^{\dagger}$ and $c_{i \alpha}$ are the 
creation and annihilation operators for an electron with spin $\alpha$ on site 
$i$; $\bm{\sigma}$ are the spin Pauli matrices. 
{ As depicted in Figs.\,\ref{model}(d) and (e), $\bm{d}_{ik}$ and 
$\bm{d}_{kj}$ are the vectors connecting the 
$i$-th and $j$-th sites to the $k$-th nearest-neighbor in common;
%
%are the vectors connecting sites $i$ and $j$,
 $t_i$ and $\lambda_i$ are the strength of hopping and spin-orbit terms.
The $\langle i j \rangle$ and $\langle \langle i j
\rangle \rangle$ refer to sums over nearest-neighbor and next-nearest-neighbor,
respectively. See 
Sec.\,I of Supplemental Material (SM)\,\cite{supplem-1}  for more details.

\section{Results and Discussions}

\subsection{Monolayer}

%%%%%FIG
\begin{figure}
\includegraphics[width=8cm]{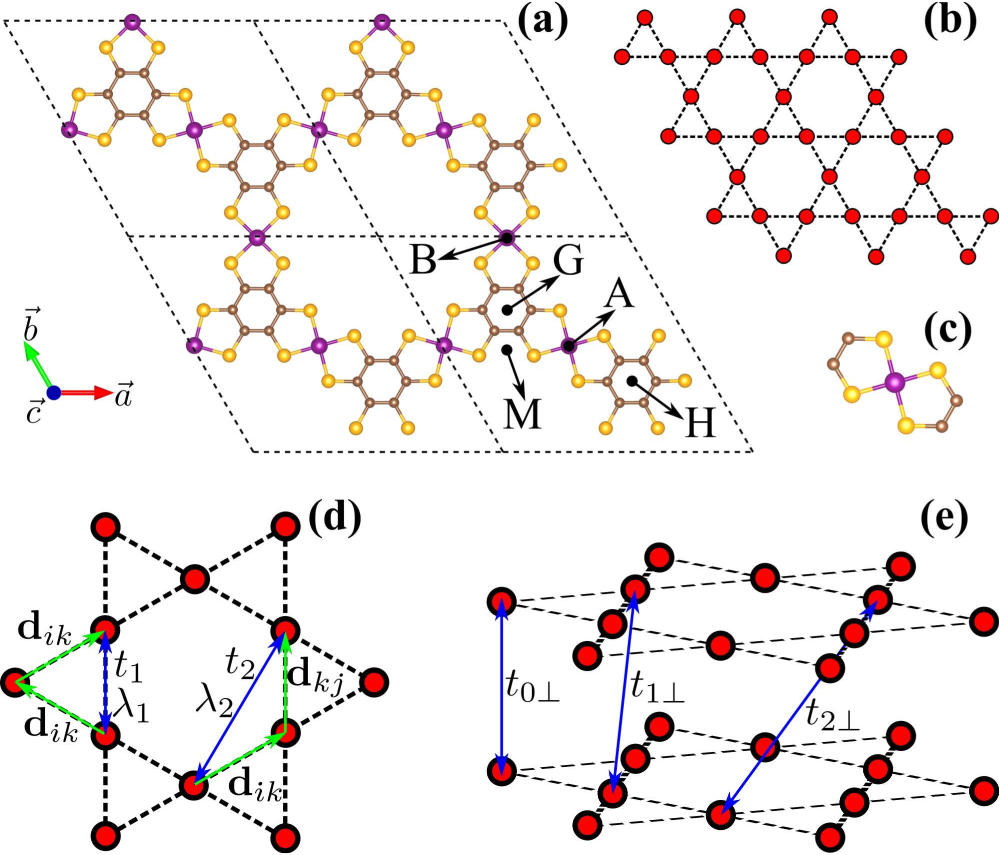}
\caption{\label{model}  (a) Atomic structure of the MOF \msc\, (b) 
representation of the kagome lattice sites (red circles) used  in the TB 
Hamiltonian, where each site are occupied by the molecule shown in panel (c).
Schematic diagram of the (d) in plane hopping and SOC parameter for nearest 
($t_1$, $\lambda_1$) and next nearest ($t_2$, $\lambda_2$) neighbors; (e) 
interlayer hopping parameter $t_0$, $t_{1\perp}$, $t_{2\perp}$. Blue lines 
indicate the coupling between two sites ($t_i$, $\lambda_i$) and green lines the 
${\bf d}_{kj}$, unity vectors of SOC term.
In panels (a) and (c) the C atoms are shown in 
brown, S in yellow and M in purple.} 
\end{figure}

The metal organic framework  of \msc\, monolayer, \msc-ML, M = Ni and Pt, 
presents a hexagonal atomic structure [Fig.\,\ref{model}(a)], which can be 
viewed as a kagome lattice [Fig.\,\ref{model}(b)], where each site is occupied 
by a (MC$_4$S$_4$) molecule [Fig.\,\ref{model}(c)]. At the equilibrium geometry, 
we found that \Ni\, presents a lattice parameter ($a$) of 14.70\,\AA, which  is 
in good agreement with recent experimental 
measurements\,\cite{kambeJACS2013}, and \fp\,  DFT 
results\,\cite{NANO2842Feng}. For {\Pt} we obtained $a = 15.06$\,\AA, as the Pt 
covalent radius is greater than Ni, which is also in agreement with recent \fp\, 
results\,\cite{CMS170Zheng}.  The electronic band structures of both MOFs 
exhibit the typical kagome energy bands above the Fermi level ($E_{\rm F}$), 
within $E_{\rm F} < E < E_{\rm F} + 0.8$\,eV. These are 
graphene-like energy bands, with a Dirac cone at the K point, 
degenerated  with a nearly flat band at the $\Gamma$ point, as shown in 
Figs.\,\ref{bands-ml}(a1) and (b1). Such degeneracies are removed 
by the SOC. In \Ni-ML we find non-trivial global energy gaps 
of 4\,meV (indirect)  and 14\,meV (direct at the K point), and a local gap of 
17\,meV, between $c2$ and $c3$ at the $\Gamma$ point, Fig.\,\ref{bands-ml}(a2). 
Those (SOC induced) energy gaps are larger in \Pt-ML, i.e. 72\,meV at 
$\Gamma$ and 60\,meV at K, as shown in Fig.\,\ref{bands-ml}(b2). Due 
to the energy dispersion of $c3$, the former is not a global gap. The larger 
energy dispersion of $c3$  can be attributed the next-nearest-neighbor 
interactions among the Pt atoms\,\cite{PRL113Zhou}. The electronic  band 
structures projected to the atomic orbitals, Figs.\,\ref{bands-ml}(a3) and (b3), 
reveal that the kagome band set of   both \msc\, systems are formed by the 
hybridization of C and S $p_z$ orbitals of the organic host, with the metal 
$d_{xz}$ and $d_{yz}$ orbitals. As will be discussed below, such a hybridization 
picture  is quite relevant for the electronic properties of the bilayer systems.

%%%%%FIG
\begin{figure}
\includegraphics[width=8cm]{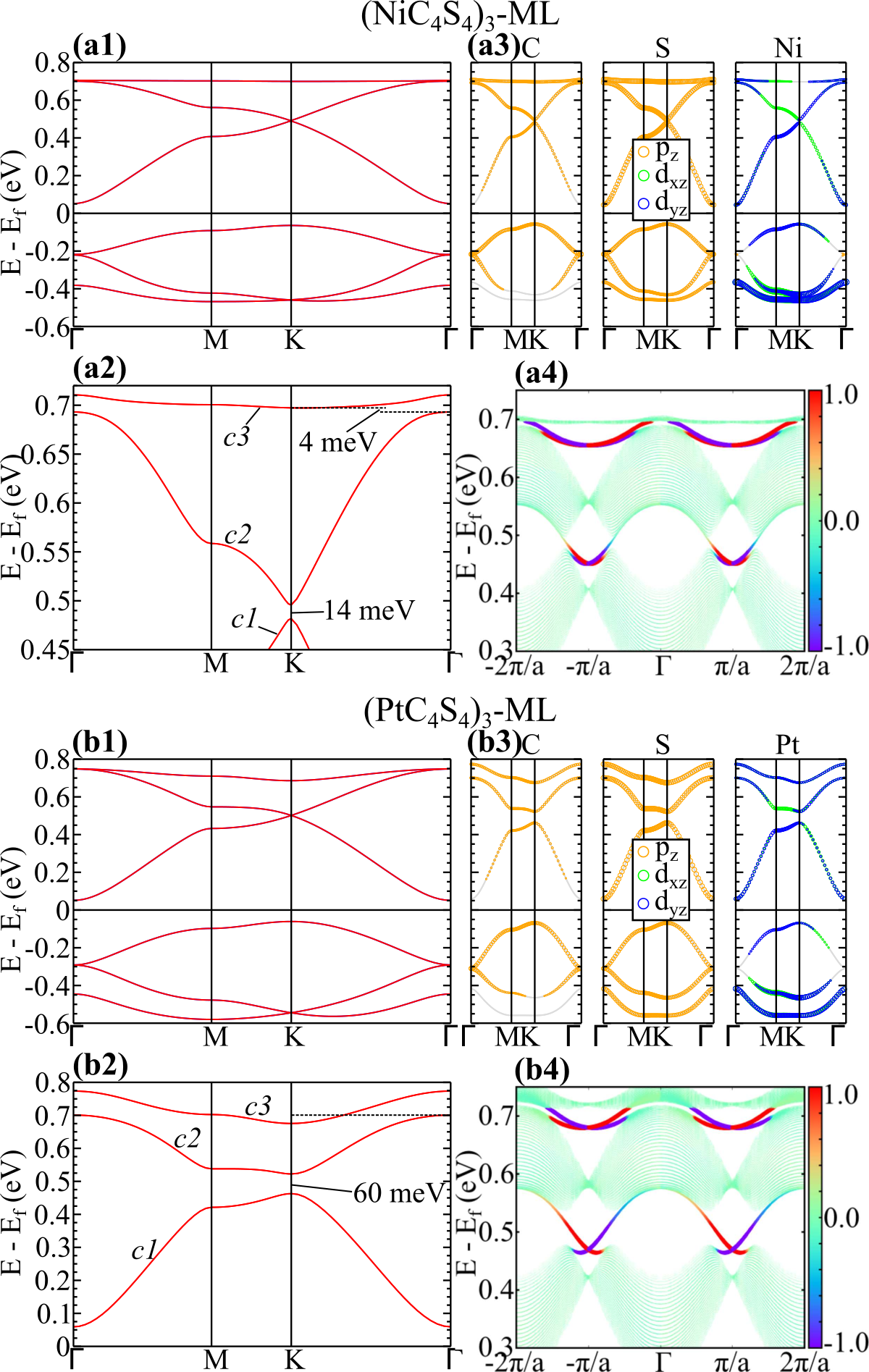}
\caption{\label{bands-ml}  First-principles electronic band 
structure of \Ni\, [(a1)-(a3)] and \Pt\, [(b1)-(b3)] MLs; without the SOC 
[(a1)-(b1)], and with the SOC [(a2)-(a3)] and [(b2)-(b3)]. The circle sizes and 
colors  [in (a3) and (b3)] indicate the contribution of each atomic orbital to 
the  band structure. Tight-binding electronic band structure [(a4)-(b4)] of 
nanoribbons of \Ni\, and \Pt\, MLs showing the chiral spin-polarized edge 
states. The color scale refers to the $\langle S_z \rangle$ component of the 
spin momentum.} 
\end{figure}

The energy gaps  induced by the SOC between $c1$ and $c2$ at K, and between $c2$ 
and $c3$ at $\Gamma$ [Figs.\,\ref{bands-ml}(a2) and (b2)] characterize the QSH 
phase of  \Ni, and \Pt. The topological phase of \Ni\,\ is well 
known\,\cite{NANO2842Feng}. Here, based on the evolution of the Wannier Charge 
Centers (WCC), we found Z$_2 = 1$ for both MOFs (Details in the 
SM\,\cite{supplem-1}). However, due to the energy dispersion of $c3$ in 
\Pt, the energy gap $c2$-$c3$  at the $\Gamma$ point is a local gap, giving rise 
to the so called  Z$_2$-metallic state\,\cite{PRB90Zhao}. Further verification 
of the QSH phase can be done by mapping the  edge states of \Ni- and \Pt-ML. 
Based on the TB approach,  we calculated the energy bands of \Ni- and \Pt-ML 
NRs. As depicted in Figs.\,\ref{bands-ml}(a4) and (b4), the formation of chiral 
spin-polarized edge states, degenerated at the TRIM,  confirms the non-trivial 
topological phases of the \Ni\, and \Pt\, MLs. We have examined the formation of 
edge states for  other edge geometries as detailed in\,\cite{supplem-1}.

\subsection{Bilayer}

In this section, based on \fp\, calculations, firstly we 
investigate the energetic stability, and the electronic properties of the 
\msc\,BL systems; and next by combining \fp\, calculations and the 
phenomenological model described below,  we provide a comprehensive  
understanding of the interlayer-electronic tuning processes mediated by an 
external electric field and interlayer separation.

The energetic stability of \msc-BL was examined by considering a set of 
different  \msc/\msc\, interface geometries, aligning sites X and Y [for X, Y 
$=$ A, B, G, H, and M, as indicated in Fig.\,\ref{model}(a)], i.e. the X site of 
one layer above the Y site of the other. In Table\,\ref{2l-binding} we show the 
averaged interlayer equilibrium distance ($d_0$), the root-mean-square deviation 
($\braket{\delta z}$) of the atomic position perpendicularly to the \msc\, 
sheet, and the BL binding energy ($E^b$). Here, we define $E^b$ as, $E^b = 
2E^{\rm (ML)} - E^{\rm (BL)}$, where $E^{\rm (ML)}$ is the total energy of an 
isolated monolayer, and $E^{\rm (BL)}$ is the total energy of the \msc-BL for a 
given  staking configuration. We found that the AA staking is the most stable 
one, with $E^b$ of 9.99 and 8.46\,meV/\AA$^2$ (69 and 62\,meV/atom) for \Ni-BL 
and \Pt-BL, respectively. Followed by the AG stacking by 0.70 and 
0.36\,meV/\AA$^2$ (4.8 and 2.6\,meV/atom). The energetic stability of those 
\msc-BLs is  ruled by vdW interactions.  It is worth noting that 
the binding strength of the \msc-BL is larger compared with other energetically 
stable 2D-vdW systems like graphene\,\cite{lebedevaPCCP2011,mostaaniPRL2015} and 
boron-nitride\,\cite{maromPRL2010,gaoPRL2015}  bilayers. There are no chemical 
bonds at the (MS$_4$C$_4$)$_3$/(MS$_4$C$_4$)$_3$ interface region, where we 
found $d_0$ of 3.64 and 3.66\,\AA\, for \Ni\, and \Pt\, BLs, and $\braket{\delta 
z} = 0.01$\,{\AA}, thus indicating that the corrugations of the \msc\, sheets 
are  negligible in the AA stacking. In contrast, the other stacking  geometries 
present $\braket{\delta z}$ between 0.1 and 0.2\,\AA.

\begin{table}[h!]
\caption{\label{2l-binding}  Bilayer binding energy $E^b = 2E^{\rm (ML)} - 
E^{(\rm BL)}$ (in meV/{\AA}$^2$), mean equilibrium interlayer  distance $d_0$ 
({\AA}) 
and root-mean-square deviation $\braket{\delta z} = \sqrt{\braket{z^2} - 
\braket{z}^2}$ ({\AA}).}
\begin{ruledtabular}
\begin{tabular}{c|cccccc}
          &  \multicolumn{3}{c}{\Ni} & \multicolumn{3}{c}{\Pt}  \\
 Staking Geometry & $E^b$ & $d_0$ & $\braket{\delta z}$ & $E^b$ & 
$d_0$ & $\braket{\delta z}$ \\
\cline{1-1} \cline{2-4} \cline{5-7}
   AA     & 9.99 & 3.64 & 0.01 & 8.46 & 3.66 & 0.01 \\
   AB     & 8.51 & 3.37 & 0.22 & 6.92 & 3.37 & 0.23 \\
   AM     & 9.26 & 3.64 & 0.08 & 7.97 & 3.60 & 0.15 \\
   AG     & 9.29 & 3.57 & 0.12 & 8.10 & 3.51 & 0.18 \\
   GH     & 8.56 & 3.35 & 0.11 & 6.92 & 3.37 & 0.13 \\
\end{tabular}
\end{ruledtabular}
\end{table}

%%%%%FIG
\begin{figure}
\includegraphics[width=8cm]{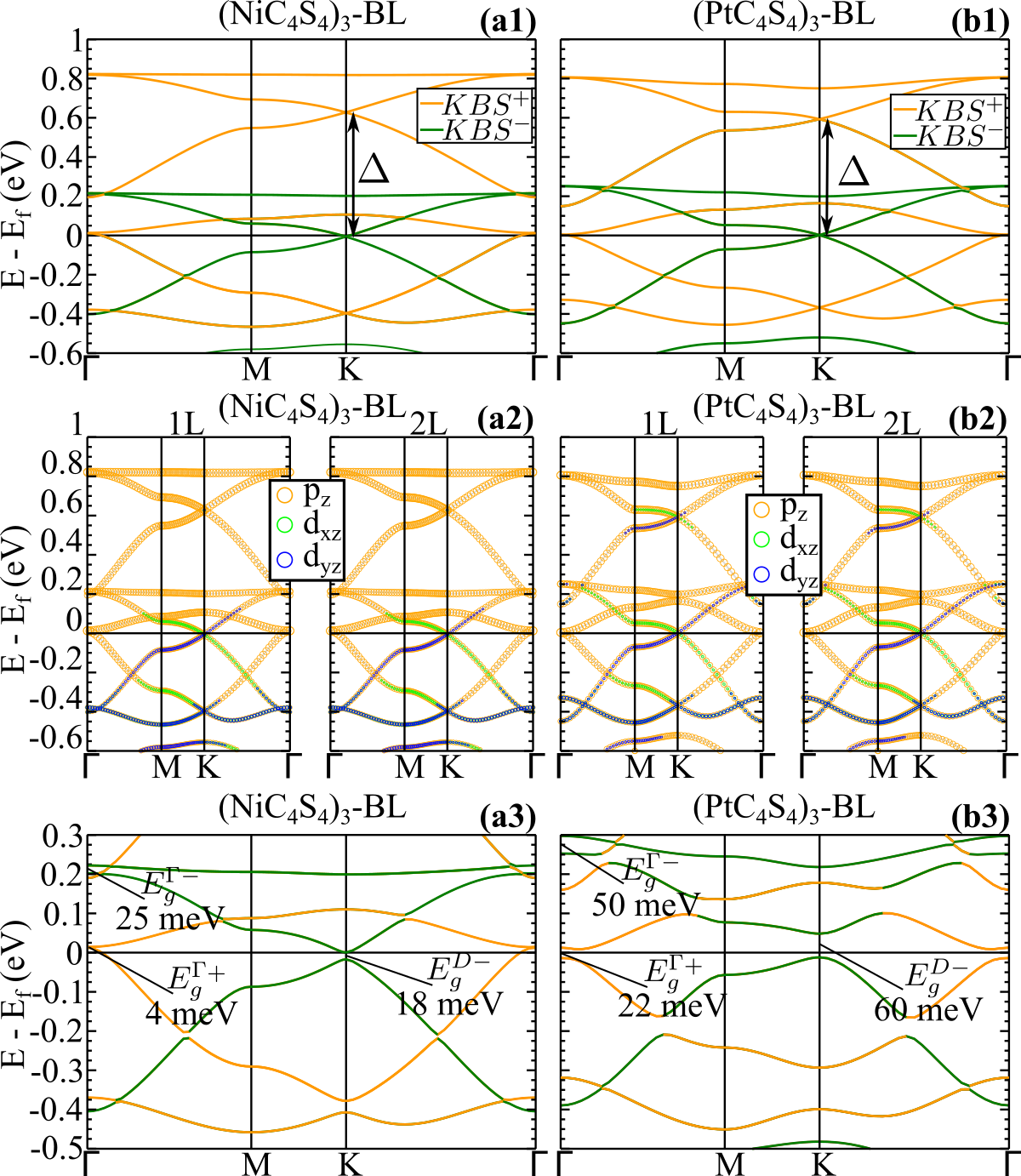}
\caption{\label{bands-bl} First-principles electronic band structure of \Ni\, (a) and \Pt\, (b) 
BLs. (a1) and (b1) band structures without SOC, where $\Delta$  indicates the 
energy  separation between the KBSs; (a2) and (b2) band structure with SOC, 
projected on the atomic orbitals of each \msc\, ML; (a3) and (b3) SOC 
induced energy gaps near the 
Fermi level.}
\end{figure}

Next we discuss the electronic properties of the energetically  most stable 
\Ni\, and \Pt\,  BLs. Initially, we will examine  the electronic band 
structure  without the SOC. The electronic structure of the BLs can be 
described as a combination of anti-bonding  (KBS$^+$) and bonding  
(KBS$^-$) kagome band sets, indicated by orange  and green  solid lines in 
Figs.\,\ref{bands-bl}(a1) and (b1). The Dirac bands of each KBSs are 
preserved, where the KBS$^+$ and KBS$^-$ are separated (in energy) by $\Delta$; 
giving rise to one Dirac point at about $E_{\rm F}+0.6$\,eV  and another 
lying on the Fermi level. Here,  $\Delta$ provide a measure  of the interlayer 
coupling between the \msc\, MLs\,\cite{supplem-1}. 
Further projected energy bands [Figs.\,\ref{bands-bl}(a2) and (b2)]  show that 
(i)  each layer exhibits the same electronic contribution on the KBS$^+$ and 
KBS$^-$, where  (ii) the energy bands are composed by $d_{xz}$ and 
$d_{yz}$ orbitals of the  transition metals (Ni and Pt) hybridized with C and S 
$p_z$ orbitals of the organic host. 

The SOC yields energy gaps at the Dirac points ($E^{\rm 
D}_{\rm g}$). For instance, in  \Ni-BL [Fig.\,\ref{bands-bl}(a3)] we find a 
energy gap of 18\,meV in KBS$^-$ ($E_{\rm g}^{\rm D-}$). This is a local energy 
gap,   due to the presence of partially occupied metallic bands near the 
$\Gamma$ point. The SOC also induces energy gaps at the $\Gamma$ point. As shown 
in Fig.\,\ref{bands-bl}(a3), we find a small local gap of 4\,meV in the KBS$^+$  
($E_{\rm g}^{\Gamma+}$) near the Fermi level, and another local gap of 25\,meV 
at $E_{\rm F}+0.2$\,eV in the KBS$^-$ ($E_{\rm g}^{\Gamma-}$). In contrast,  
\Pt\, BL presents a global gap of 22\,meV at the Fermi level ($E_{\rm 
g}^{\Gamma+}$), followed by $E_{\rm g}^{\rm D-}$ of 60\,meV, and a local gap of 
50\,meV at the $\Gamma$ point ($E_{\rm g}^{\Gamma-}$) [Fig.\ref{bands-bl}(b3)]. 
As will be discussed below, those  energy gaps induced by the SOC will dictate 
the formation of topologically protected edge states in the \msc-BLs.

To model the DFT results presented above, we propose a
phenomenological Hamiltonian to describe the interaction between layers.
Assuming the mirror symmetry of the AA stacking, the Hamiltonian reads
\begin{equation} H_s = h_{3 \times 3}({\bf{k}}) \otimes \, \tau_0 + 
\frac{\Delta}{2}\, \mathbb{I}_{3 \times 3} \otimes \tau_x, 
\end{equation} 
where, $h_{3 \times 3}({\bf k})$, represents the Hamiltonian of each monolayer 
separately, diagonal on the base $\{\ket{\#L;n,{\bf k}}\}$ ($n = 1, 2, 3$ bands, 
$\# = 1, 2$ layers), which gives the kagome band dispersions; $\tau_j$ 
($j=0,\,x,\,y,\,z$) are the Pauli matrix in the layer space, and $\Delta /2$ the 
coupling term between the layers. In  this model, each layer will interact 
forming the highest energy (anti-bonding, $\ket{+}$) and the lowest energy 
(bonding, $\ket{-}$)  KBSs,  energetically separated by $\Delta$.  In 
this case, the Dirac bands at the Fermi level are given by the bonding 
KBSs, green solid lines in Fig.\,\ref{bands-bl}(a1) and (b1). The mirror 
symmetry imposes  that $|\braket{\#L|\pm}|^2=1/2$, for \#~=~1, 2. 

The mirror symmetry can be suppressed upon  the interaction of the \msc-BLs with 
a solid surface, or due to the presence of an external electric  field 
perpendicular to the \msc\, layer. The latter can be expressed by  adding a 
potential difference between the layers in $H_s$, 
\begin{equation} H = H_s - 
\varepsilon\, \mathbb{I}_{3 \times 3} \otimes \tau_z. 
\end{equation} 
Here, the potential difference due to only the external electric field 
($E^{\rm ext}$) will be 
$\varepsilon = (d/2) E^{\rm ext}$, but the charge rearrangement at the 
\msc/\msc\, interface can reduce this potential difference such that, 
$\varepsilon = \sigma E^{\rm ext}$. Further discussion on the proposed model can 
be found in the Supplemental Material\,\cite{supplem-1}, Sec.\,II. Therefore, in 
this model the contribution of each layer to an given state is $E^{\rm ext}$ 
dependent. 

Initially, the effect of external electric field  was studied based on the 
\fp\,  approach.  In Figs.\,\ref{bands-bl-efield}(a1) and (b1) we present the 
electronic band structures of the \Pt\, and \Ni\, BLs for  $E^{\rm ext}=0$.
The mirror symmetry is fulfilled and both layer contributes equally for 
each state. The size of red circles is  proportional to the layer contribution 
to each  state, $|\braket{\# L|n,{\bf k}}|^2$. By turning on the 
external electric field ($E^{\rm ext}\neq 0$), there is an unbalance on the 
charge density distribution between the MLs,  Figs.\,\ref{bands-bl-efield}(a2) 
and (b2);  followed by an increase on the energy separation between the kagome 
bands, $\Delta=0.63\rightarrow0.94$\,eV as the electric field module increase 
from $0.0\rightarrow0.2$\,eV/{\AA} in  \Ni-BL. In contrast, such an increase of 
$\Delta$, as a function of the external field, is almost negligible in \Pt-BL. 
For the electric field module increasing from $0.0\rightarrow0.5$\,eV/{\AA},  
the separation between the kagome bands changes by less than 0.03\,eV 
($\Delta=0.59\rightarrow0.61$\,eV). 

%%%%%%FIG
\begin{figure}[!htb]
\includegraphics[width=8cm]{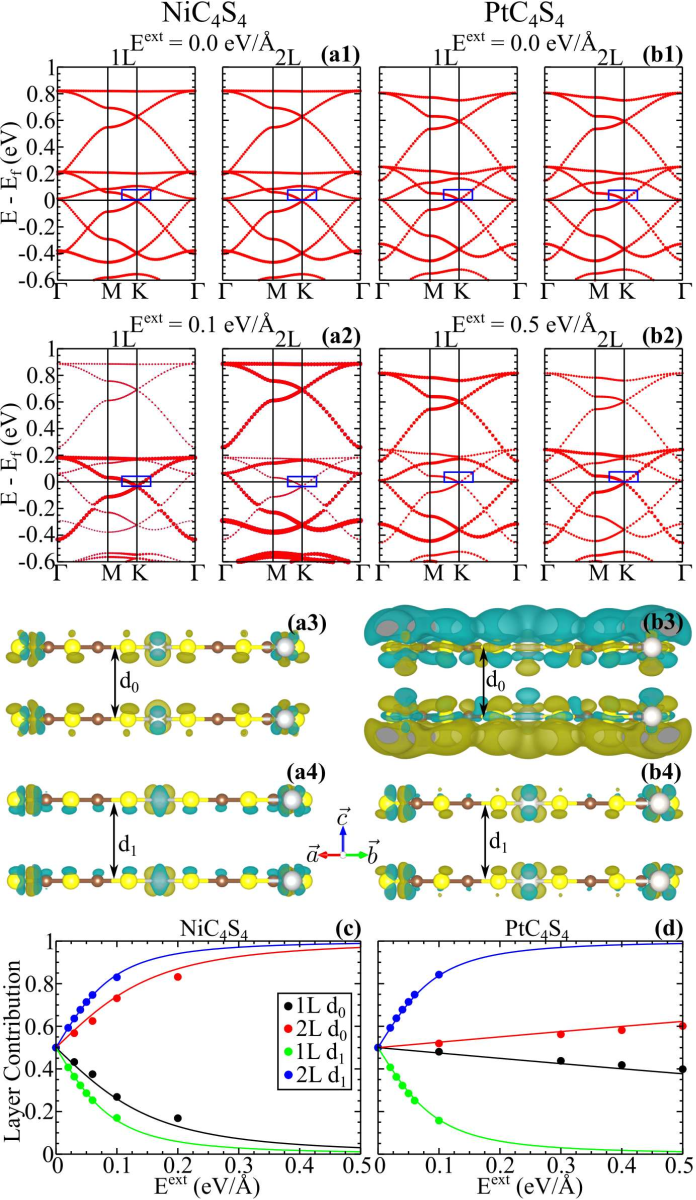}
\caption{\label{bands-bl-efield}  First-principles electronic structure of \Ni\,(a) and 
\Pt\, (b) BLs. (a1) and (b1) Energy bands projected on the different MLs, 
$|\braket{\#L|n,{\bm k}}|^2$ for $\#L$= 1L and 2L with $E^{\rm ext}=0$; (a2) and (b2) with $E^{\rm ext} \neq 0$. Total charge density difference, 
$\Delta\rho$, due to an external field of 
0.06\,eV/\AA\ and with distance $d_0=3.6$\,\AA\, (a3) and (b3); 
$\Delta\rho$ with distance $d_1=3.9$\,\AA\, 
(a4) and (b4). The isosurfaces in (a3)-(a4) and (b3)-(b4) is 
$2\times10^6$\,e/{\AA}$^3$, where regions in yellow (blue) indicate an 
electron gain 
(loss). In (c) and (d) we show the contribution of each layer for the KBS 
$|\braket{1L|-}|^2$ (black and green), and $|\braket{2L|-}|^2$ (red and blue) 
for $d_0$ and $d_1$; the results obtained through the 
phenomenological model (\fp\, DFT) are indicated 
by the solid lines (circles). }
\end{figure}

The dependence of $\Delta$  with  the external electric field can be understood 
by analyzing the changes on the total charge density ($\Delta\rho$) as a 
function of $E^{\rm ext}$ and the interlayer distance $d$. For a given 
value of $d$, we can define $\Delta\rho$ as,
\begin{equation}
 \Delta\rho = \rho(E^{\rm ext}) - \rho(0),
\end{equation}
where $\rho(E^{\rm ext})$ and $\rho(0)$ represent the total charge densities of 
the \msc\ BL at $E^{\rm ext}\neq0$ and $E^{\rm ext}=0$, respectively. Our 
results of $\Delta\rho$ for the \Ni\, and \Pt\, BLs show that, (i)  at the 
equilibrium geometry ($d_0=3.6$\,\AA), there is no   charge transfer between the 
\Ni\, MLs [Fig.\,\ref{bands-bl-efield}(a3)]; in contrast (ii)  a net charge 
transfer takes place between the \Pt\, MLs [Fig.\,\ref{bands-bl-efield}(b3)]. 
Such a net charge transfer gives rise to an intrinsic local electric field which 
can be written as, $E^{\rm loc} = -\alpha E^{\rm ext}$; reducing the potential 
difference between the \Pt\, MLs, in agreement with the small changes on the 
energy separation between the kagome bands, $\Delta$. By 
increasing the interlayer distance, for instance $d_0\rightarrow d_1=3.9$\,\AA, 
we found that (i) the electronic interaction between the  \msc\, MLs reduces, as 
well as the coupling term $\Delta$. We found $\Delta=0.41$\,eV (\fp\, 
calculations) for both \msc\,BLs; and (ii)  there is  a reduction on the net 
charge transfer between the MLs due to the external electric field, as depicted 
in Figs.\,\ref{bands-bl-efield}(a4) and (b4) for {\Ni} and {\Pt}, respectively.

As shown in Figs.\,\ref{bands-bl-efield}(a2) and (b2), the layer contribution on 
the KBSs can be controlled by  an external electric field. Here we will consider 
the electronic states around the Dirac point near 
the the Fermi level, indicated by (blue) rectangles in 
Figs.\,\ref{bands-bl-efield}(a1)-(a2) and \ref{bands-bl-efield}(b1)-(b2). The 
calculated partial charge densities within those rectangles, $|\braket{1L|-}|^2$ 
and   $|\braket{2L|-}|^2$, are shown in Figs.\,\ref{bands-bl-efield}(c) and (d) 
for $E^{\rm ext}$ from 0 to 0.5\,eV/\AA. Our \fp\, results are indicated 
by colored circles, and solid lines indicate the ones  obtained by using  the 
phenomenological model. At $E^{\rm ext}=0$ we  have $|\braket{1L|-}|^2 = 
|\braket{2L|-}|^2 = 0.5$, {\it i. e.}  both layers present the same electronic 
contribution as the mirror symmetry is fulfilled. For lower values of 
$E^{\rm ext}$,  $\varepsilon \ll \Delta /2$,  
the electronic contribution of each layer exhibits a linear behaviour, where the 
tangent modulus is $\sigma/\Delta$\,\cite{supplem-1}. The separation of the 
partial charge densities between the MLs is  strengthened for larger interlayer 
distances. For instance, at the equilibrium geometry, $d_0=3.6$\,\AA, we 
find $|\braket{1L|-}|^2=0.27$ and $|\braket{2L|-}|^2=0.73$, which corresponds to 
a charge density  separation ratio ($\eta$),  
$$
\eta=\frac{|\braket{1L|-}|^2}{|\braket{2L|-}|^2}
$$
of 0.37 for $E^{\rm ext}=0.1$\,eV/\AA\, in \Ni-BL;  increasing $d$ to 3.9\,\AA, 
the charge density separation increases,  $\eta=0.20$ for the same value of 
$E^{\rm ext}$. On the other hand, the net charge transfers between the \Pt\, MLs 
result in   $\varepsilon \ll \Delta / 2$ for a greater range of $E^{\rm ext}$, 
giving rise to a linear response of the layer contribution, even for $E^{\rm 
ext} = 0.5$\,eV/\AA, black and red lines in Fig.\,\ref{bands-bl-efield}(d). 
Indeed, for $d_0=3.6$\,\AA\,  the  charge density separation is very small,  we 
find $|\braket{1L|-}|^2=0.40$ and $|\braket{2L|-}|^2=0.60$,    $\eta=0.67$ for 
$E^{\rm ext} = 0.50$\,eV/\AA. On the other hand, increasing the interlayer 
distance to $d = 3.9$\,{\AA}, the charge transfer is suppressed 
[Fig.\,\ref{bands-bl-efield}(b4)], and  we find $\eta=0.19$  for $E^{\rm 
ext}=0.10$\,eV/{\AA}, which is  practically the result obtained in \Ni-BL.

It is worth noting that (i) by inverting the $E^{\rm ext}$ 
direction, the layer localization also inverts ($1L \leftrightarrow 2L$), and 
(ii) in the present scenario the  charge density separation in 
\msc\ BLs is  ruled by the suppression of the mirror symmetry. Here we have 
considered the suppression of the mirror symmetry through an external electric 
field, but the same behavior is expected in other cases, e.g. the presence of a 
substrate. In the next section we discuss the bilayers ribbons and the location 
of the topologically protected edge  states, by the breaking of the mirror 
symmetry.

%%%%%FIG
\begin{figure}[!htb]
\centering
\includegraphics[width=8cm]{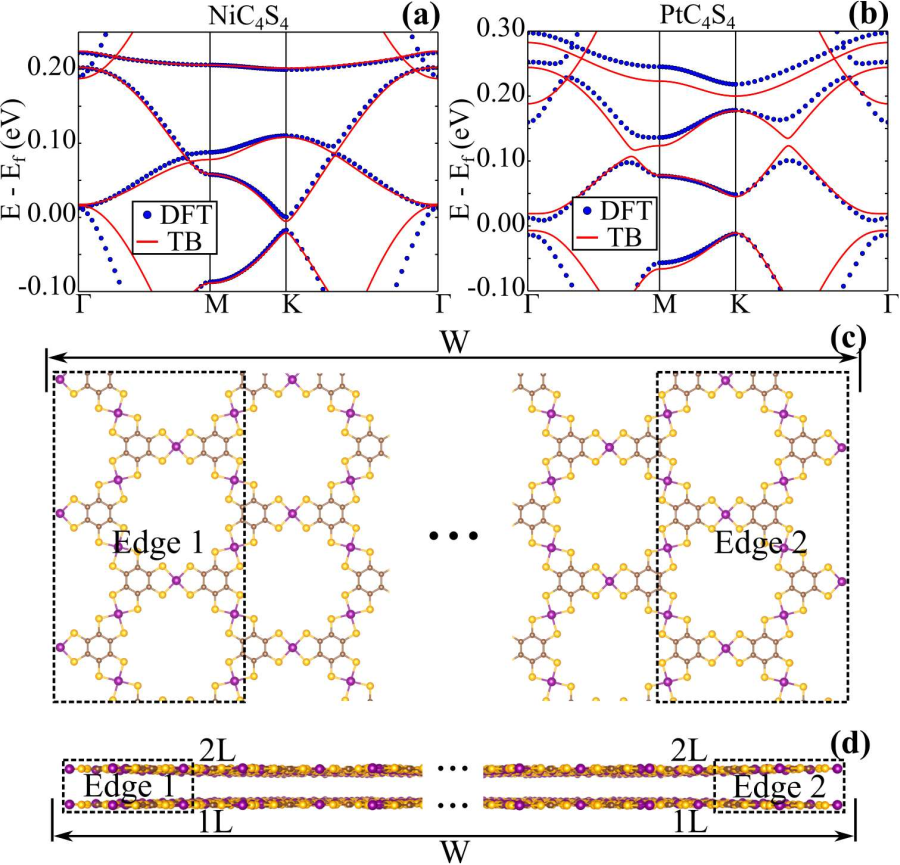}	
\caption{\label{tb-2l-bulk}  Electronic band structure of 
\Ni\,(a), and \Pt\,(b) bilayers, obtained through \fp\, (blue circle), and TB 
(red line) calculations
 for $d_{\rm LL}=3.6$\,\AA\, and including the SOC.
Bilayer nanoribbon geometry, top-view (c) and side-view (d).}
\end{figure}

\subsection{Bilayer Nanoribbon}

In this section we will discuss the edge states of BL nanoribbons, 
in order to provide a more complete picture of the electronic properties of the 
\msc-BLs.  Here, the electronic band structure of \Ni- and \Pt-BLs, obtained 
through \fp\, calculations,  was fitted within the TB approach considering the 
intralayer and interlayer hoppings, and    two orbitals (A and B) per site of 
the kagome bilayer-lattice (details in Sec.\,I of the SM). As shown in 
Figs.\,\ref{tb-2l-bulk}(a) and (b), the energy dispersions obtained through the 
TB Hamiltonian (red lines) present a reasonably well correspondence with the 
ones obtained by the \fp\, calculations approach (blue circles),  where the main 
features of the band structure are well described.

    Similarly to the monolayers, the bilayers also have a $\mathbb{Z}_2 = 1$ 
topological invariant. However, here it shows a $\mathbb{Z}_2 = 1$ for each set 
of orthogonal subspaces. For the mirror symmetry case, these are the bonding 
(KBS$^+$) and anti-bonding (KBS$^-$) states. For a finite $E^{ext}$ it is still 
possible to define two orthogonal sets [see Supplemental 
Material\,\cite{supplem-1}], which are similar to the KBS$^\pm$ states. 
Consequently, in the following we find two sets of edge states, one for each 
orthogonal subspace.

In  order to identify these
topologically protected edge states we have considered nanoribbon widths (W) of  
$\sim 51$ and $\sim 52$\,nm for {\Ni} and {\Pt} BLs, Figs.\,\ref{tb-2l-bulk}(c) 
and (d).

%%%%%FIG
\begin{figure}[!htb]
\includegraphics[width=8cm]{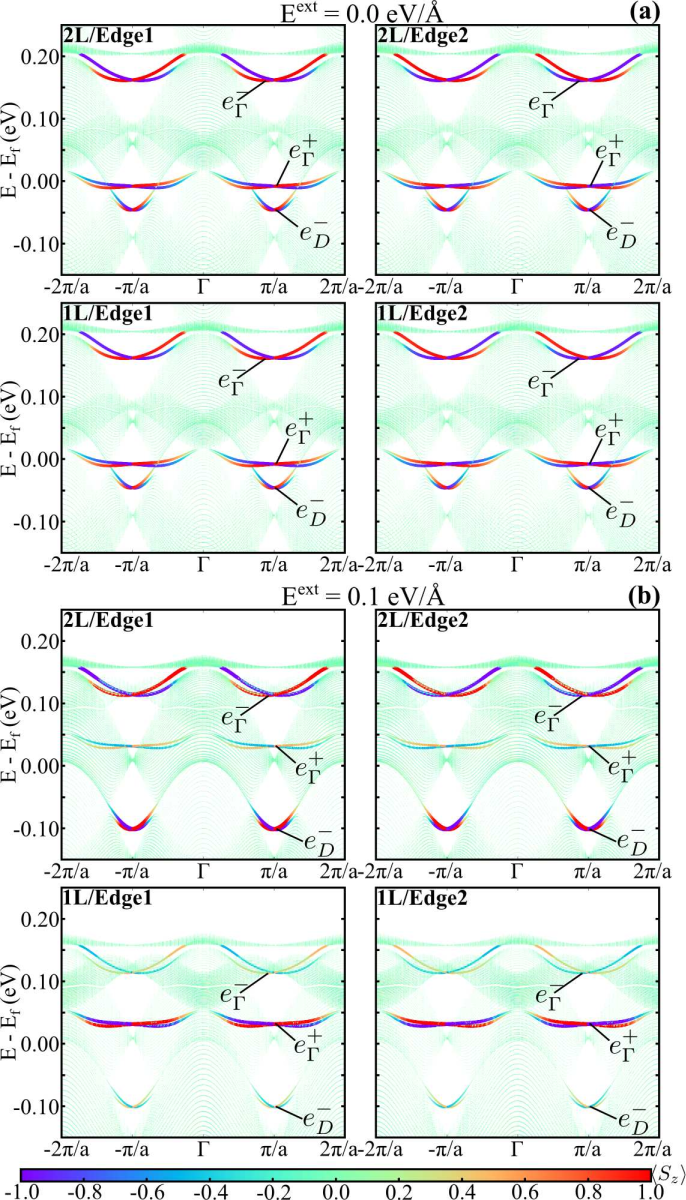}
\caption{\label{ni-ribbon} Spin polarized electronic band structure of the {\Ni}-BL NR, projected on the edge atoms [Edge 1 and Edge 2 indicated in Figs.\ref{tb-2l-bulk}(c) and (d)] for $E^{\rm ext}=0$ (a), and $E^{\rm  ext}=0.1$\,eV/\AA\, (b).  The color map are the $\braket{S_z}$, and the size of  projection is proportional to $|\braket{{\# L,\rm Edge}|n, {\bf k }}|^2$.}
\end{figure}

In Figs.\,\ref{ni-ribbon} and \ref{pt-ribbon} we present the spin-polarized 
energy bands projected on the edge atoms  indicated in 
Figs.\,\ref{tb-2l-bulk}(c) and (d). The   formation of chiral edge 
states, degenerated at the TRIM, confirms  the topological phases of the \Ni\, 
and \Pt\, BLs. Figures\,\ref{ni-ribbon}(a) and \ref{pt-ribbon}(a) show  three 
sets of metallic edge states   near the Fermi level, {\it viz.}: $e^-_{\Gamma}$, 
$e^+_{\Gamma}$, and  $e^-_{\rm D}$, degenerated at the TRIM k = $\pi$/a and 
$-\pi$/a. Those metallic bands  come from the (non-trivial) energy gaps induced 
by the SOC, between (i) the kagome bands at the $\Gamma$ point, 
$E^{\Gamma-}_{\rm g}$, and  $E^{\Gamma+}_{\rm g}$, and (ii) the Dirac states at 
the K point, $E^{\rm D-}_{\rm g}$.   In \Ni-BL, the  energy gaps   (i) and (ii) 
[Fig.\,\ref{bands-bl}(a3)] are not global, however, the formation of spin 
polarized  chiral edge states supports the so called Z$_2$-metallic phase. 
In contrast, \Pt-BL 
presents a (large) global energy gap of  22\,meV ($E_{\rm g}^{\Gamma+}$) at 
the Fermi level, Fig.\,\ref{bands-bl}(b3). Thus, in order to acess the QSH 
phase in \Pt-BL, it is not necessary any external doping to place the 
Fermi level in the non trivial energy gap.

%%%%%%FIG
\begin{figure}
\includegraphics[width=8cm]{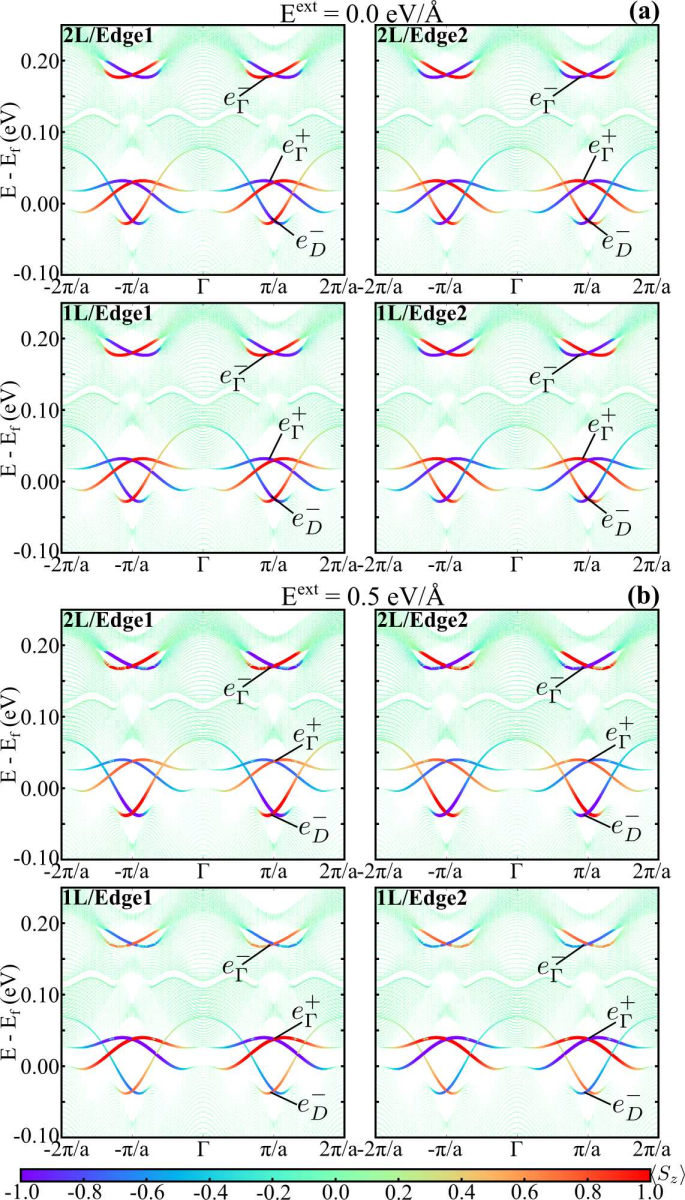}
\caption{\label{pt-ribbon}   Spin polarized electronic band structure of the {\Pt}-BL NR, projected on the edge atoms [Edge 1 and Edge 2 indicated Figs.\ref{tb-2l-bulk}(c) and (d)] for $E^{\rm ext}=0$ (a), and $E^{\rm ext}=0.5$\,eV/\AA\, (b).  The color map are the $\braket{S_z}$, and the size of projection is proportional to $|\braket{{\#L,\rm Edge}|n,{\bf k }}|^2$.}
\end{figure}

Edge states localized at opposite sides of the same \msc\, 
ML present opposite chiralities; while edge states of different MLs 
present the same chirality and the same electronic density of states. The latter 
is a consequence of the mirror symmetry in the BL systems. As we have shown 
in the previous section,  the mirror symmetry can be removed by turning on an 
external electric field perpendicularly to the \msc\, layers. Here we will 
focus on the interlayer separation of the topological edge states mediated by 
an external electric field.  

The effect of $E^{\rm ext}$ on the localization of the edge states are depicted 
in Figs.\,\ref{ni-ribbon}(b) and \ref{pt-ribbon}(b). In the former 
diagram we present the localization of the edge states in the \Ni-BL NR
upon an external field of $0.1$\,eV/\AA. The edge states $e^-_{\Gamma}$ and 
$e^-_{\rm D}$ are mostly localized on one ML (1L), while  $e^+_{\Gamma}$ 
lies on the 
other (2L).

Similarly to what we have done in the BL systems, we can define a separation 
ratio for the edge states  ($\eta_e$). In order to compare with the $\eta$ 
calculated for  \msc-BLs, we will consider the (layer) projection of 
the edge states $e^-_{\rm D}$,
$$
\eta_e = \frac{|\braket{1L|e^-_{\rm D}}|^2}{|\braket{2L|e^-_{\rm D}}|^2}. 
$$
We find $\eta_e=0.39$ for  ($E^{\rm ext}=0.1$\,eV/\AA), which is 
practically the same result obtained in \Ni-BL, at the same external field. 
As shown in Fig.\,\ref{pt-ribbon}(b), such a  charge density separation also 
takes place in the \Pt-BL NR, however is not so effective.  This is a 
consequence of the net charge transfer verified in \Pt-BL, as discussed above. 
Indeed, here we find  $|\braket{1L|e^-_{\rm D}}|^2=0.39$, and 
$|\braket{2L|e^-_{\rm D}}|^2=0.61$, giving rise to $\eta_e=0.64$, for $E^{\rm 
ext}=0.5$\,eV/\AA; which is  practically the same value obtained in \Pt-BL,  
$\eta=0.67$. 

On the other hand, as shown in Fig.\,\ref{bands-bl-efield},  the separation of 
the partial charge densities in \msc-BLs can be tuned by changing the vertical 
distance between the \msc\ sheets. For instance, $\eta$ reduces from 0.67 to 
0.19 by increasing the interlayer distance from 3.6 to 3.9\,\AA\, in \Pt-BL.  
Edge states in \Pt-BL NR present the same behavior. By increasing the interlayer 
distance to  $d_1=3.9$\,\AA, the layer separation of the edge states is 
strengthened, where we find   $|\braket{1L|e^-_{\rm D}}|^2=0.18$, and 
$|\braket{2L|e^-_{\rm D}}|^2=0.82$, $\eta_e=0.22$, for $E^{\rm ext} = 0.1$\,eV/\AA .
That is, in addition to the 
external electric field, the interlayer distance is another degree of freedom 
which allow us  to control the localization of the topologically protected edge 
states in \msc\ BL nanoribbons.   It is worth to mention that such a control on 
the interlayer distance, between the organic layers, can be done through  the 
current  pillaring processes in MOFs\,\cite{eubankJACS2011,sunJPorousMat2016}.

\section{Conclusion}

Based on \fp\, calculations and  tight-binding model, we show that the energetic 
stability of the  \msc-BLs, M = Ni and Pt, is ruled by vdW interactions, being 
the AA stacking the most stable one. The electronic structure of the \msc-BLs 
is characterized by the formation of bonding and anti-bonding KBSs; where the 
energy gaps in the KBSs, induced by the SOC, give rise to QSH  or 
Z$_2$-metallic state in \msc-BLs. Their  topologically non-trivial nature 
was  identified through the formation of  chiral spin-polarized edge states. By 
considering  a phenomenological model, combined with \fp\, calculations, we 
present a very comprehensive picture of the electronic properties upon 
the presence of an external electric field. In this case, the electronic 
contributions from each  ML,   to the formation of the  bonding and anti-bonding 
KBSs, are no longer symmetric, with the bonding and anti-bonding KBSs localized 
in different MLs. We find  that the chiral edge states follow the same 
 pattern, and thus the localization of the 
topologically protected edge states in \msc-BLs can be tuned by an external 
electric field. Our findings are not restricted to the \Ni\, and \Pt\ BLs.
We can infer that such a tuning process
 will also take place  in other mirror symmetric vdW metal-organic BLs characterized by 
a superposition of the kagome bands.

\section{ACKNOWLEDGMENTS}

The authors acknowledge   financial   support   from   the Brazilian  agencies  
CNPq, and FAPEMIG, and the  CENAPAD-SP for computer time.

\bibliography{bib}% Produces the bibliography via BibTeX.

%%%%%%%%%%%%%%%%%%%%%%%%%%%%%%%%%%%%%%%%%%%%%%%%%%%%%%%%%%%%%%%%%%%%%%%%%%%%%%%
%%%%%%%%%%%%%%%%%%%%%%%%% SUPPLEMENTAL MATERIAL %%%%%%%%%%%%%%%%%%%%%%%%%%%%%%%
%%%%%%%%%%%%%%%%%%%%%%%%%%%%%%%%%%%%%%%%%%%%%%%%%%%%%%%%%%%%%%%%%%%%%%%%%%%%%%%

\onecolumngrid
\clearpage

\begin{center}
\textbf{\large Supplemental Materials: Tuning the topological states in metal-organic bilayers}
\end{center}

%%%%%%%%%% Prefix a "S" to all equations, figures, tables
\setcounter{equation}{0}
\setcounter{figure}{0}
\setcounter{table}{0}
\setcounter{page}{1}
\makeatletter
\renewcommand{\theequation}{S\arabic{equation}}
\renewcommand{\thefigure}{S\arabic{figure}}
\renewcommand{\thetable}{S\Roman{table}}
\renewcommand{\bibnumfmt}[1]{[S#1]}
\renewcommand{\citenumfont}[1]{S#1}
%%%%%%%%%%%
%\makeatother

 In the present Supplemental Material, we provide details of the 
TB 
model applied for the {\msc}-ML/BL systems, Section~I;  describe the 
phenomenological Hamiltonian for mirror symmetric bilayers and the effect of 
breaking this mirror symmetry, Section~II; and analyse the Rashba spin-orbit 
contribution to the topological phases of the \msc-ML and BL, Section~III. In 
this latter section, based on the TB model and first-principles calculations,  
we  show that the Rashba  spin-orbit contribution can be safely neglected in the 
present study.

\section{Kagome Lattice Tight-Binding Model}
		The real-space tight-binding Hamiltonian of kagome-hexagonal lattice in the presence of intrinsic spin-orbit coupling (SOC) [S1, S2] can be written as
	\begin{equation}
	H_{TB} = H_0 + H_{SO} \label{tb}
	\end{equation}
where each term is given by
	$$
	H_0 = t_1 \sum_{\langle i j\rangle; \alpha} c_{i \alpha}^{\dagger} c_{j \alpha} 
	+ t_2 \sum_{\langle \langle i j \rangle \rangle; \alpha} c_{i \alpha}^{\dagger} 
	c_{j \alpha} ;
	$$
	\begin{eqnarray*}
	H_{SO} = i\, {\lambda}_{1} \sum_{ \langle i j \rangle} c_{i}^{\dagger} 
	\bm{\sigma} \cdot (\bm{d}_{kj} \times \bm{d}_{ik}) c_j + \\ i\, {\lambda}_{2} 
	\sum_{\langle \langle i j \rangle \rangle} c_{i}^{\dagger} \bm{\sigma} \cdot 
	(\bm{d}_{kj} \times \bm{d}_{ik}) c_j ;
	\end{eqnarray*}
here, $c_{i \alpha}^{\dagger}$ and $c_{i \alpha}$ are the creation and annihilation operators for an electron with spin $\alpha$ on site $i$; $\bm{\sigma}$ the spin Pauli matrices, 
{\color{black} $\bm{d}_{ik}$ and $\bm{d}_{kj}$ the unity vector connecting the 
$i$-th and j-$th$ sites to the $k$-th nearest-neighbor in common 
[Fig.\,\ref{hopping}(a)], such that $\bm{d}_{kj} \times \bm{d}_{ik} = \pm 
(\sqrt{3}/2) \hat{\bm e}_z$ (where the proportionality constant $\sqrt{3}/2$ is 
absorbed in $\lambda_i$)},
 and $t_i$, $\lambda_i$ the strength of hopping and spin-orbit terms. The 
$\langle i j \rangle$ and $\langle \langle i j \rangle \rangle$ refer to 
nearest-neighbor and next nearest-neighbor summation, respectively. It worth to 
mention that we have done DFT {\fp} calculation on the {\msc} monolayers (ML) 
with perpendicular external electric field up to $0.5$\,eV/{\AA}, in which we do 
not observe any Rashba SOC effect on the band structure. Therefore in our TB 
model we have taken the Rashba SOC as null [{\color{black} more details on the 
Rashba SOC term in the last section}].
		
		The kagome lattice is given by a hexagonal lattice with 3 atoms as base [see main text, Fig.1]. For this lattice we can identify the nearest ($t_1$) and next nearest ($t_2$) neighbors in plane hopping, as represented in Fig.\ref{hopping}(a). By the formation of bilayer systems in addition to the in plane hopping, interlayer hopping are considered as represented in FIg.\,\ref{hopping}(b).  
%%%%%%FIG
\begin{figure}[h!]
\centering
\includegraphics[width=13cm]{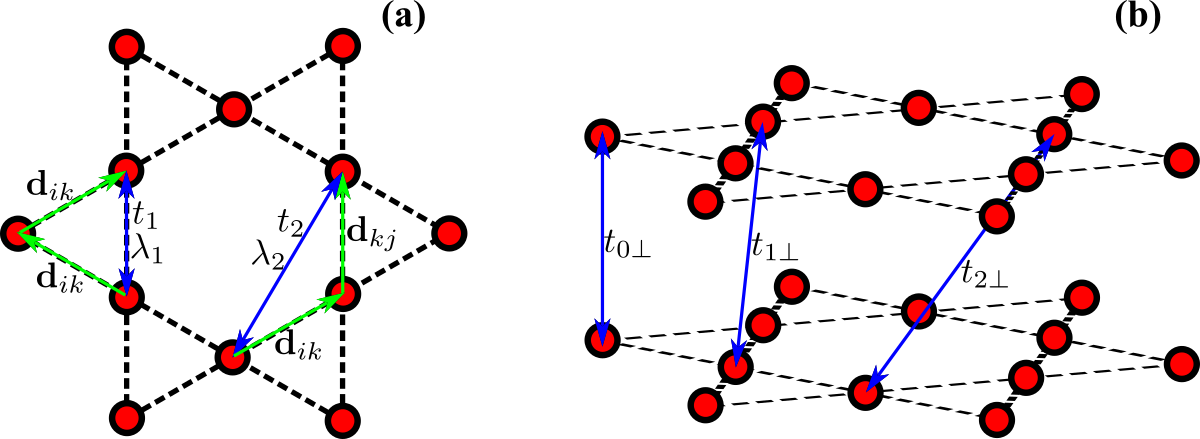}
\caption{\label{hopping} (a) In plane hopping and SOC parameter for nearest ($t_1$, $\lambda_1$) and next nearest ($t_2$, $\lambda_2$) neighbors; (b) Interlayer hopping parameter $t_0$, $t_{1\perp}$, $t_{2\perp}$. Blue lines indicate the coupling between two sites ($t_i$, $\lambda_i$) and green lines the ${\bf d}_{kj}$, unity vectors of SOC term.}
\end{figure}

%\section{Topological Characterization: {\msc}-ML}
		For {\msc}-ML (M= Ni, Pt) we find a good TB description of the system by fitting the {\fp} band structure with the parameters show in TABLE\,\ref{tb-1l-parameter}, considering a single orbital per kagome site. We can see that the single orbital TB Hamiltonian describe well the kagome band set (KBS) dispersion [Fig.\,\ref{tb-1l-band}(a1) and (b1)] obtained by DFT {\fp} calculations [see main file methodology to DFT description].

%%%%%%TAB
\begin{table}[!ht]
\centering
\caption{\label{tb-1l-parameter} Monolayer TB parameters in meV.}
%\begin{ruledtabular}
\begin{tabular}{c|cc}
\hline \hline
Parameter & {\Ni} & {\Pt} \\
\hline
$E_0$            &  550.0  &  575.0  \\
$t_{1}$          &  -38.0  &  -40.0  \\
$t_{2}$          &   -2.0  &   -3.0  \\
$\lambda_{1}$    &   -1.8  &   -4.7  \\
$\lambda_{2}$    &   -0.2  &   -1.7  \\
\hline \hline
\end{tabular}
%\end{ruledtabular}
\end{table}

		Within the TB hamiltonian we can track the evolution of Wannier Charge Center (WCC) of an effective 1D system. Where the $\mathbb{Z}_2$ topological invariant number is given by WCC evolution through half pumping cycle
\begin{equation}
	\mathbb{Z}_2 = \sum_{\alpha} \left[ \bar{x}^{I}_{\alpha}(\pi) - \bar{x}^{II}_{\alpha}(\pi) \right] - \sum_{\alpha} \left[ \bar{x}^{I}_{\alpha}(0) - \bar{x}^{II}_{\alpha}(0) \right]\bmod 2 \label{z2},
\end{equation}
where $I$, $II$ are for Kramers pairs, and ${\alpha}$ are the index of occupied states in terms of pairs [S3, S4]. One graphical way to visualize the $\mathbb{Z}_2$, is to draw a reference arbitrary vertical line from $k_y = 0$ to $k_y = \pi$ on the WCC evolution, where even (odd) number of crossing with the reference line determine the topological trivial (non-trivial) characteristic of the system [S4].

%%%%%%FIG
\begin{figure}[h!]
\centering
\includegraphics[width=17cm]{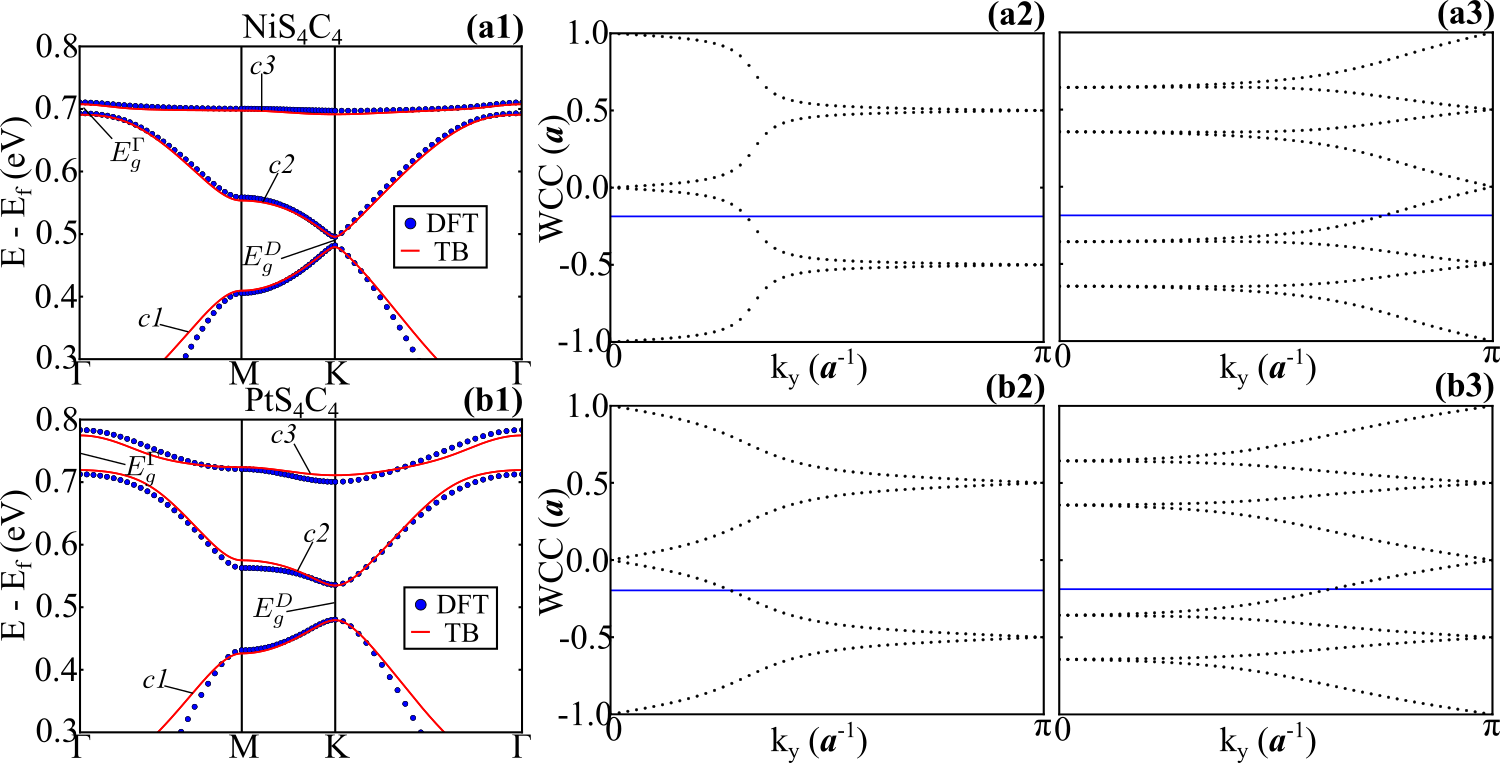}
\caption{\label{tb-1l-band} {\Ni} (a) and {\Pt} (b) band structure and Wannier Charge Center (WCC) evolution. TB (red solid line) and DFT (blue circle) band structure (a1) and (b1); WCC for the occupation up to $E_g^D$ gap (a2) and (b2); WCC for the occupation up to $E_g^{\Gamma}$ gap (a3) and (b3). }
\end{figure}
		By considering the occupation of {\msc}-ML up to $E_g^D$, namely band $c1$ fully occupied and $c2$, $c3$ unoccupied [Fig.\,\ref{tb-1l-band}(a1) and (b1)], we can calculate the WCC of the system. As show in Fig.\,\ref{tb-1l-band}(a2) and (b2) for {\Ni} and {\Pt], respectively, the reference line (blue solid line) cross the evolution line one time, therefore $\mathbb{Z}_2 = 1$ which characterizes this gap as topologically non-trivial. In a same way, for the occupation up to $E_g^{\Gamma}$, namely $c1$ and $c2$ band fully occupied [Fig.\,\ref{tb-1l-band}(a1) and (b1)], the reference line cross the evolution line one time as show in Fig.\,\ref{tb-1l-band}(a3) and (b3) for {\Ni} and {\Pt}, respectively. We can conclude that both gaps $E_g^{D}$ and $E_g^{\Gamma}$ are topologically non-trivial, and making a lateral interface with a trivial material is expected a chiral edge state.

		Next we have constructed nanoribbons of {\msc} with width (W) of $\sim 50$\,nm, with three different edge geometries, Fig.\,\ref{tb-1l-edge}(a), (b) and (c). Here are observed the formation of chiral edge states within the bulk SOC induced gap energy $E_g^D$ ($\sim 0.5$\,eV) and $E_g^{\Gamma}$ ($\sim 0.7$). From Fig.\,\ref{tb-1l-edge}(a) we can see that the ribbon geometry does not have inversion symmetry as we have two different type of edges, {\it i. e.} one terminated with a chain of atoms (Edge 1) and other by triangles (Edge 2). Within this asymmetric nanoribbon, each edge has a chiral momentum-spin locked states, degenerated at the time reversal invariant momenta (TRIM) in $ k = \pm \pi / a$, but with different Fermi velocities. On the other hand, for the inversion symmetric nanoribbons, Fig.\,\ref{tb-1l-edge}(b) and (c), the chiral states of opposite edges has the same Fermi velocity.

%%%%%%FIG
\begin{figure}[h!]
\centering
\includegraphics[width=17cm]{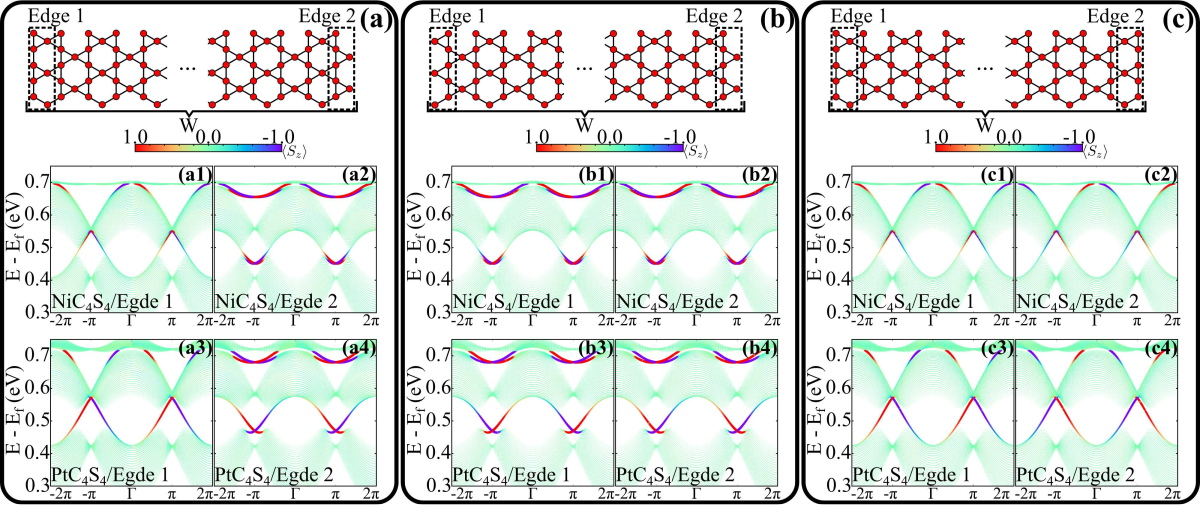}
\caption{\label{tb-1l-edge} {\msc}-ML nanoribbon with three different edge geometries (a), (b), (c). Edge projected band structure, with the color scale indicating $\braket{S_z}$; {\Ni}/Edge 1 in (a1), (b1), (c1); {\Ni}/Edge 2 in (a2), (b2), (c2); {\Pt}/Edge 1 in (a3), (b3), (c3); {\Pt}/Edge 2 in (a4), (b4), (c4).}
\end{figure}

		For the {\msc} bilayer TB model we considered two mirror symmetric kagome lattices. In addition to this doubled system we have considered two orbitals per kagome site (A and B), as the monolayer KBS below the Fermi level [see Fig. 2(a1) and (b1) in the main file] become relevant near the bilayer Fermi level. To discriminate the SOC and hopping strength for different orbitals we use the notation $t_i^{(XY)}$, meaning the $i$th neighbor hopping between the orbital X and Y (with X,Y $=$ A,B), as presented in TABLE\,\ref{tb-2l-parameter}. This parameters well described the main characteristics of the {\fp} band structure as show in Fig.\,5(c) and (d) of the main file.

%%%%%%TAB
\begin{table}[!ht]
\centering
\caption{\label{tb-2l-parameter} Bilayer TB parameters in meV.}
%\begin{ruledtabular}
\begin{tabular}{c|cc||c|cc}
\hline \hline
Parameter & {\Ni} & {\Pt} & Parameter & {\Ni} & {\Pt} \\
\hline
$E_0^{(A)}$            &  360.0  &  369.0 & $E_0^{(B)}$            & -506.0  & -488.0 \\

$t_0$                  &  291.0  &  291.0 & $t_{1}^{(AA)}$         &  -46.7  &  -47.5 \\

$t_{2}^{(AA)}$         &   -1.0  &   -5.0 & $t_{1}^{(BB)}$         &  -45.6  &  -45.6  \\

$t_{2}^{(BB)}$         &    7.6  &   14.0 & $t_{1}^{(AB)}$         &    0.0  &    0.0  \\

$t_{2}^{(AB)}$         &    0.0  &    0.0 & ${\lambda}_{1}^{(AA)}$ &   -1.8  &   -4.7  \\

${\lambda}_{2}^{(AA)}$ &   -0.2  &   -1.7 & ${\lambda}_{1}^{(BB)}$ &   -1.9  &   -6.0  \\

${\lambda}_{2}^{(BB)}$ &   -1.8  &   -4.5 & ${\lambda}_{1}^{(AB)}$ &    0.0  &    0.0  \\

${\lambda}_{2}^{(AB)}$ &    0.0  &    0.0 & $t_{1\perp}^{(AA)}$    &   -9.0  &   -7.0  \\

$t_{2\perp}^{(AA)}$    &    0.0  &    0.0 & $t_{1\perp}^{(BB)}$    &  -20.0  &  -20.0  \\

$t_{2\perp}^{(BB)}$    &    0.0  &    0.0 & $t_{1\perp}^{(AB)}$    &    2.0  &    8.0  \\

$t_{2\perp}^{(AB)}$    &   -2.0  &   -2.0 &                        &         &         \\
\hline \hline
\end{tabular}
%\end{ruledtabular}
\end{table}

		For the mirror symmetric {\msc}-BL Hamiltonian, one can break the mirror symmetry by adding a potential difference between the layers through the on site energy of each layer orbital, {\it i. e.} making $E_0 \rightarrow E_0 \pm \varepsilon/2$, with $+$ ($-$) sign for the upper (lower) layer. This potential difference can be due to an external electric field ($E^{\rm ext}$). Taking {\Ni}-BL as an example, we can see that in the absence of external electric field [Fig.\,\ref{tb-efield}(a)] the system is mirror symmetric and each layer contribute equally to each state. On the other hand, by breaking the mirror symmetry the contribution of each layer to an given state are $E^{ext}$ dependent, as show the color map fin Fig.\,\ref{tb-efield}(b) and (c). We introduce a phenomenological model to explain the effects of this mirror symmetry breaking, as discussed in the next section.
		
%%%%%%FIG
\begin{figure}[h!]
\centering
\includegraphics[width=12cm]{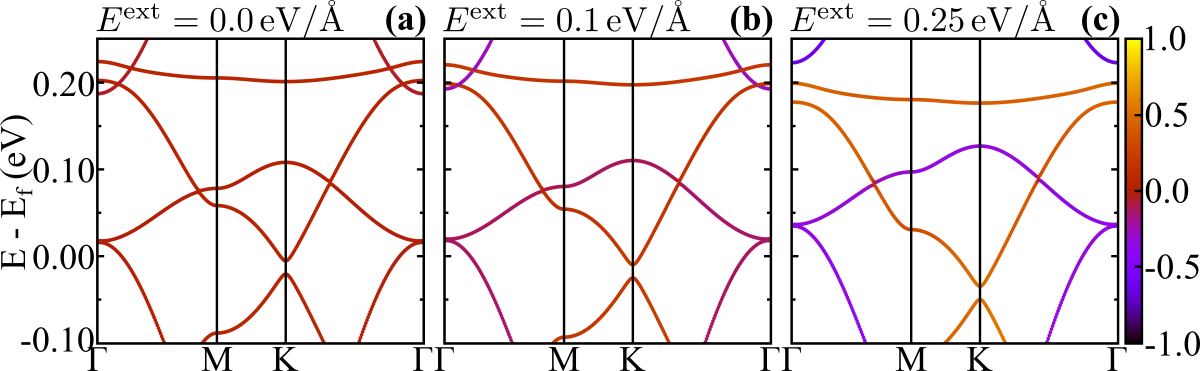}
\caption{\label{tb-efield} {\Ni}-BL band structure with external electric field of $0.0$ (a), $0.1$ (b) and $0.25$\,eV/{\AA} (c). The color map indicates $|\braket{2L|n, {\bf k}}|^2 - |\braket{1L|n, {\bf k}}|^2$}
\label{FigMirror}
\end{figure}

\section{Phenomenological Model: {\msc}-BL}

Here we propose a phenomenological model to describe the interaction between 
the layers for the AA staking configuration. Firstly, knowing the Hamiltonian of 
each monolayer separately $h_{3 \times 3}({\bf k})$, diagonal on the base 
$\{\ket{\#L;n,{\bf k}}\}$ ($n = 1, 2, 3$ bands, $\# = 1, 2$ layers), which gives 
each kagome band set (KBS) dispersions. Therefore, the general Hamiltonian for 
the two identical layers in the AA staking interacting with each other, and with 
an external electric field perpendicular to the layers is
		\begin{equation}
		H = h_{3 \times 3}({\bf{k}}) \otimes \, \tau_0 + 
\frac{\Delta}{2}\, \mathbb{I}_{3 \times 3} \otimes \tau_x - \varepsilon\, 
\mathbb{I}_{3 \times 3} \otimes \tau_z, 
		\end{equation}
where $\tau_j$ ($j = x, y, z$) are the Pauli matrix in the layers space, $\Delta 
/2$ is the coupling between the layers and $\varepsilon$ the potential energy 
associated with the electric field. Here we omit the spin degree of freedom, since time-reversal is preserved. Also $\varepsilon$ describe the asymmetry of the 
two layers upon presence of external electric field. If, on the presence of 
external electric field, the two layers does not exchange electrons, the potential 
energy between the layers is just $V = d_{LL}\,E^{ext}$ (for $E^{ext}$ in 
eV/{\AA}), and defining $\varepsilon = \sigma \, E^{ext}$, we have $\sigma = 
d_{LL} / 2$. On the other hand, if the presence of external electric field make 
the layers exchange electrons, the potential energy between the layers is 
weakened by the presence of a induced local electric field $E^{loc} = - \alpha 
\, E^{ext}$, therefore in this case $V = d_{LL}(1 - \alpha)\,E^{ext}$, and for 
$\varepsilon = \sigma E^{ext}$ $\rightarrow$ $\sigma = d_{LL}(1 - \alpha)/2$. 
The eigenvalues and eigenstates of this interacting layer model are given by
		\begin{equation}
		E_{\pm}^{(n)} = h^{(n)}({\bm k}) \pm \sqrt{\varepsilon^2 + 
\left( \frac{\Delta}{2} \right)^2 },\label{eigval}
		\end{equation}
		\begin{equation}
		\ket{\pm;n,{\bf k}}  = A_{\pm} \left\{ \ket{1L;n,{\bf k}} \pm 
B_{\pm} \ket{2L;n,{\bf k}} \right\}\label{state1}
		\end{equation}
with
		\begin{eqnarray*}
		A_{\pm} = \left\{ 1 + {B_{\pm}}^2 \right\}^{-1/2}\; ; \\ \\ \; 
B_{\pm} = \left[ \frac{\sqrt{\varepsilon^2 + (\Delta / 2)^2} \pm \varepsilon 
}{(\Delta/2)} \right],\label{state2}
		\end{eqnarray*}
and the band index $n = 1, 2, 3$. Therefore we can write the energy separation 
between each KBS as $E_+ - E_- = 2 \sqrt{ \varepsilon^2 + (\Delta/2)^2}$.

If the electric field is absent ($\varepsilon = 0$) the 
Hamiltonian commutes with the mirror symmetry, such that the eigenvalues and 
eigenstates are
		\begin{equation}
		E_{\pm}^{(n)} = h^{(n)}({\bm k}) \pm 
\left|{\frac{\Delta}{2}}\right|,\label{mirror-eig}
		\end{equation}
		\begin{equation}
		 \ket{\pm;n,{\bf k}} = \frac{1}{\sqrt{2}} \left( \ket{1L;n,{\bf 
k}} \pm \ket{2L;n,{\bf k}} \right)\label{mirror-ket}.
		\end{equation}	
These solutions show us that: (i) for $\Delta \neq 0$ the band structure is 
composed by two KBS (symmetric $\ket{+}$ and anti-symmetric $\ket{-}$) separated 
in energy by $\Delta$ ($ = E_+ - E_-$), and as long as the mirror symmetry 
is present, each state is an equal linear combination of the state from each layer 
($|\braket{\#L|\pm}|^2=1/2$); on the other hand (ii) for $\Delta = 0$ the system present two 
fold degenerated KBS (regardless of spin). 

In contrast, for $\varepsilon 
\neq 0$ the contribution of each layer to a given band is $E^{ext}$ dependent, 
been $|A_{\pm}|^2$ and $1 - |A_{\pm}|^2$ for the layer 1L and 2L, respectively. For 
$\varepsilon \ll \Delta / 2$ these contributions are linear with 
$E^{ext}$,
		\begin{eqnarray}
		|\braket{1L|\pm}|^2 \approx \frac{1}{2} \pm \frac{\sigma}{\Delta} 
E^{ext} ;\label{Alin} \\
		|\braket{2L|\pm}|^2 \approx \frac{1}{2} \mp \frac{\sigma}{\Delta} 
E^{ext}.
		\end{eqnarray}

In Section I we have seen that each monolayer returns the topological invariant $\mathbb{Z}_2 = 1$. Interestingly, the bilayer system also has $\mathbb{Z}_2 = 1$. As shown above, for a mirror symmetric bilayer ($\varepsilon = 0$), we find KBS states formed by bonding and anti-bonding orthogonal subspaces, which are energy split by $\Delta$ [see Eqs.\,\eqref{mirror-eig}-\eqref{mirror-ket}]. These are eigenstates of the mirror symmetry with eigenvalues $m_z = \pm 1$. Within each subspace, the topological invariant is $\mathbb{Z}^{m_z}_2=1$. For a finite $\varepsilon$ the mirror symmetry is broken. Nonetheless, it is still possible to label the KBS states by the orthogonal subspaces defined by the eigenstates of $\tilde{M} = (\Delta/2 \,\tau_x - \varepsilon\, \tau_z)/\sqrt{\varepsilon^2+(\Delta/2)^2}$ [see Eq.~\eqref{state1}], with eigenvalues $\tilde{m}_z = \pm 1$. This simply generalizes the mirror symmetry and yields $\mathbb{Z}^{\tilde{m}_z}_2=1$ for each subspace. Notice that for $\varepsilon \ll \Delta$, $\tilde{M}$ reduces to the mirror symmetry operator, while for $\varepsilon \gg \Delta$ it labels the top and bottom layers, as shown in Fig.~\ref{FigMirror}.

%%%%%%FIG
\begin{figure}[h!]
    \centering
    \includegraphics[width=17cm]{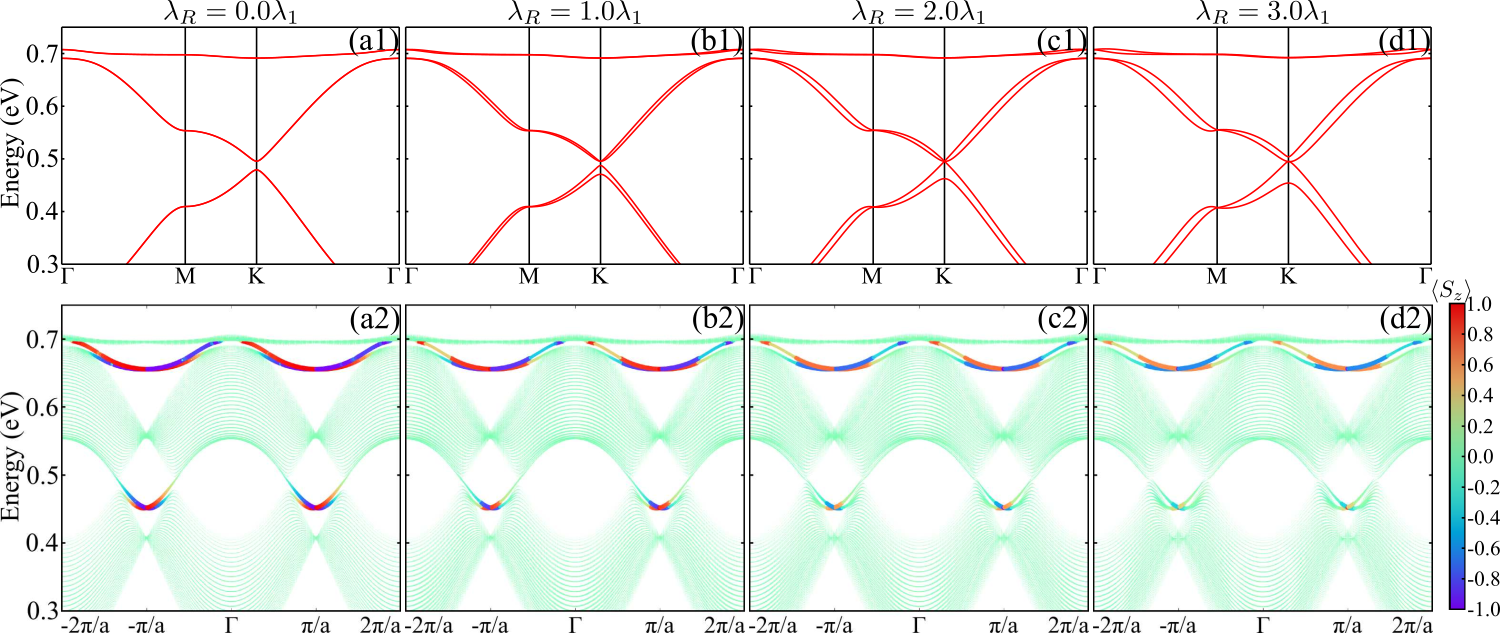}
    \caption{\label{ml-rashba} Electronic band structures, obtained by the  TB 
        approach for the kagome \msc-ML [Eq.\,\eqref{tb-rashba}], for different 
        Rashba SOC strength ($\lambda_R$); bulk states (a1)-(d1), and nanoribbon edge 
        states  (a2)-(d2) projected on the $\langle S_z \rangle$ component of the 
        spin-polarization.}
\end{figure}

\section{Rashba Spin-Orbit Coupling}

The contribution of the Rashba SOC on the electronic band structure of the \msc-ML and -BL systems is defined by the term $H_R$ added to the tight-binding Hamiltonian
\begin{equation}
    H_{TB} = H_0 + H_{SO} + H_R, \label{tb-rashba}
\end{equation}
where $H_0$ and $H_{SO}$ are given in Eq.\eqref{tb}, and
	$$
	H_R = i \lambda_R \sum_{\langle ij \rangle} c_{i}^{\dagger} \hat{\bm e}_z \cdot ({\bm \sigma} \times {\bf d}_{ij})c_{j},
	$$
with $d_{ij}$ the unitary vector connecting the $i$-th to the $j$-th site. The 
effect of Rashba SOC in the low energy Dirac dispersion of graphene is well 
known;  the energy gap induced by intrinsic spin-orbit coupling vanish for large 
Rashba SOC contribution, characterized by a spin-splitting of the energy 
bands\,[S5, S6]. Here, we find a somewhat similar picture in the 
\msc-ML and -BL. In Figs.\,\ref{ml-rashba}(a1)-(d1)  we present the energy bands 
of \msc-ML as a function of the  the strength of the Rashba SOC ($\lambda_R$) in 
$H_{TB}$,  $\lambda_R=0\rightarrow3\lambda_1$, where  we can identify the 
spin-splitting due to the Rashba SOC for $\lambda_R \ge \lambda_1$. Further 
consequences on the edges states of \msc\, nanorribons, upon the inclusion of  
$H_R$, are shown in Figs.\,\ref{ml-rashba}(a2)-(d2). The edge states are 
preserved, however,  their chiral spin polarizations fade out by increasing the 
Rashba SOC contribution ($\lambda_R=0\rightarrow3\lambda_1$). It is interesting 
to note that   even for large Rashba contribution, 
e.g. $\lambda_R=2\lambda_1$  in Fig.\,\ref{ml-rashba}(c), the (spin) 
chirality of the edge states has been preserved. In the sequence, based on 
\fp\, calculations, we will show that the contributions of the Rashba SOC are 
quite small in the present \msc-ML and BL systems.

%%%%%%FIG
\begin{figure}[h!]
\centering
\includegraphics[width=13cm]{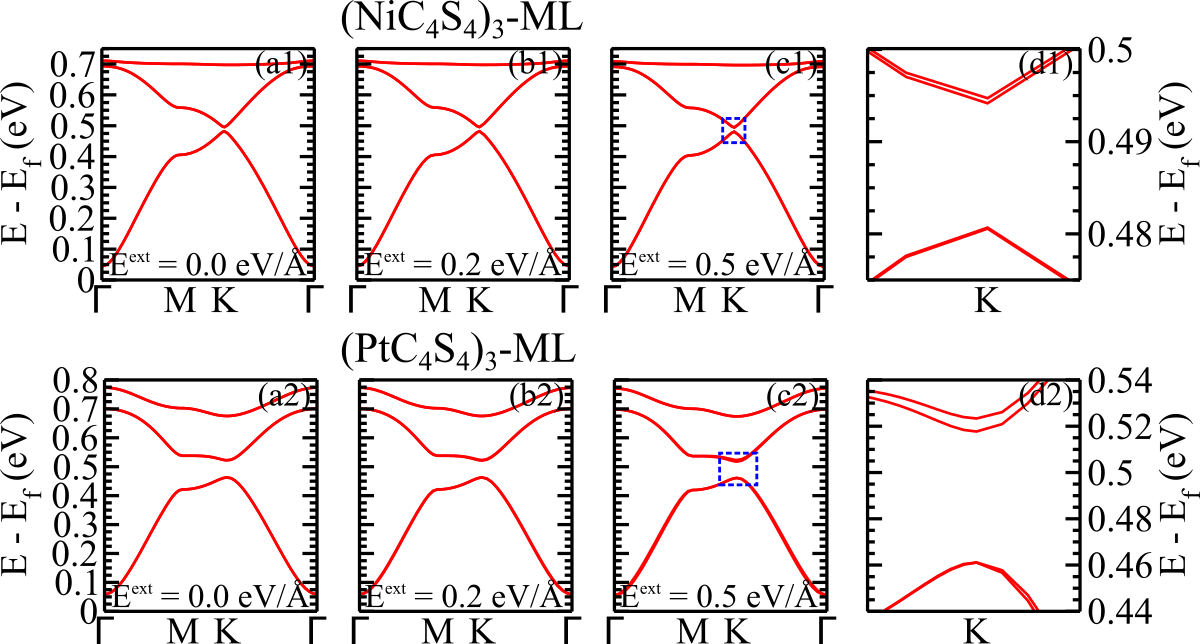}
\caption{\label{ml-fp} First-principles electronic band structure as a function 
of the external electric field E$^{\rm ext}$. (a1)-(c1) {\Ni}-ML and 
(a2)-(c2) {\Pt}-ML. In (d1) and (d2) we evidenciate the blue dashed retangles of (c1) and (c2) respectively.}
\end{figure}

An external electric field perpendicular to the \msc\, layers  suppresses the 
mirror  of the system, and thus promoting the Rashba SOC effects. In order to 
provide a quantitative measure  of such a contribution on the \Ni\, and \Pt-ML 
and -BL systems, we perform a set of \fp\, calculations of the electronic band 
structures of those \msc\, systems upon the presence of an external electric 
field ($E^{\rm ext}$). In Figs.\,\ref{ml-fp}(a1)-(c1) and \ref{ml-fp}(a2)-(c2) 
we present the electronic band structure of \Ni- and \Pt-MLs for $E^{\rm 
ext}=0\rightarrow0.5$\,eV/\AA; where we find that the effect of Rashba 
SOC is much smaller compared with the one of the  intrinsic SOC, up to 
$E^{\rm ext} = 0.5$\,eV/{\AA}. For instance, the {\Ni}-ML exhibits a  
a  spin-splitting  of about $0.5$\,meV  for $E^{\rm ext} = 0.5$\,eV/{\AA}, 
which is small in comparison with the  (intrinsic SOC) energy gap of $14$\,meV 
(less than $4\%$), Fig.\,\ref{ml-fp}(d1). Similarly, in \Pt-ML, we found a spin-splitting of 
$\sim$5\,meV and an energy gap of 60\,meV due  to the intrinsic SOC, Fig.\,\ref{ml-fp}(d2).

%%%%%%FIG
\begin{figure}[h!]
\centering
\includegraphics[width=17cm]{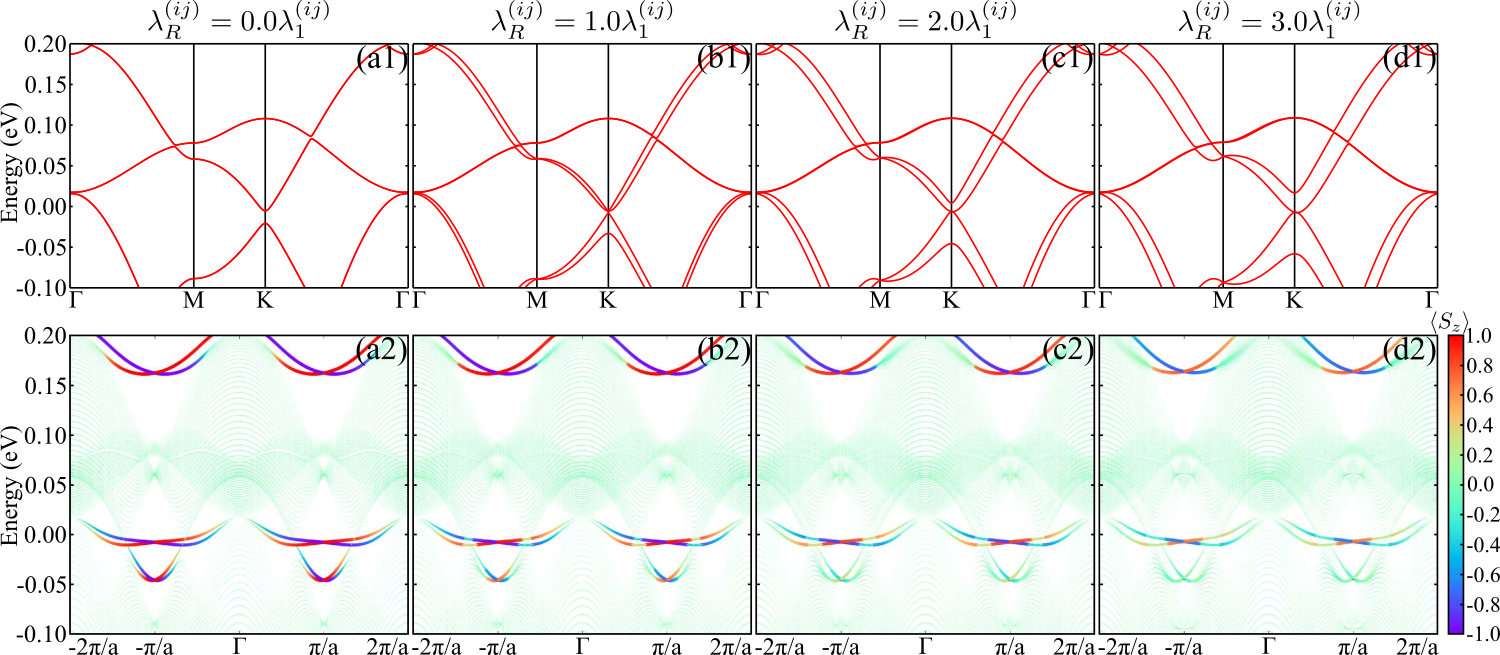}
\caption{\label{bl-rashba} Electronic band structures, obtained by the  TB 
approach for the kagome \msc-BL [Eq.\,\eqref{tb-rashba}], for different 
Rashba SOC strength ($\lambda_R$); bulk states (a1)-(d1), and nanoribbon edge 
states  (a2)-(d2) projected on the $\langle S_z \rangle$ component of the 
spin-polarization. $i,\,j=A,\,B$.}
\end{figure}

Following the same scheme, based on the TB model [Eq.\,\eqref{tb-rashba}], in 
Figs.\,\ref{bl-rashba}(a1)-(d1) we present the electronic band structures of the 
\msc-BL, as a function of the strength of the $H_R$ contribution ($\lambda_R$), 
and the respective spin-polarized edge states, Figs.\,\ref{bl-rashba}(a2)-(d2). 
Similarly to what we found in the single layer systems, (i) the spin-chirality 
of the edges fades out for larger values of $\lambda_R$ in comparison with 
$\lambda_1$, however  (ii) even for $\lambda_R=2\lambda_1$, the chiral character 
of the (edge) energy bands has been maintained. In the sequence, we performed 
\fp\, calculations of the electronic band structures of \Ni-BL and \Pt-BLs as a 
function of the external electric field. Our results are summarized in  
Fig.\,\ref{bl-fp}, where we  show the evolution of the electronic band 
structures of  \Ni-BL (a1)-(c1) and \Pt-BL (a2)-(c2). For both systems, we found 
that the spin-splitting due to the Rashba SOC about ten times smaller than the 
energy gap induced by the intrinsic SOC, Figs.\,\ref{bl-fp}(d1) and (d2). 

Therefore, we can infer that the effect of Rashba SOC is small, compared with 
the other contributions,  and can be dismissed in the (present) \msc\, monolayer 
and bilayer systems for $E^{\rm ext}$ within the studied range.

%%%%%%FIG
\begin{figure}[h!]
\centering
\includegraphics[width=13cm]{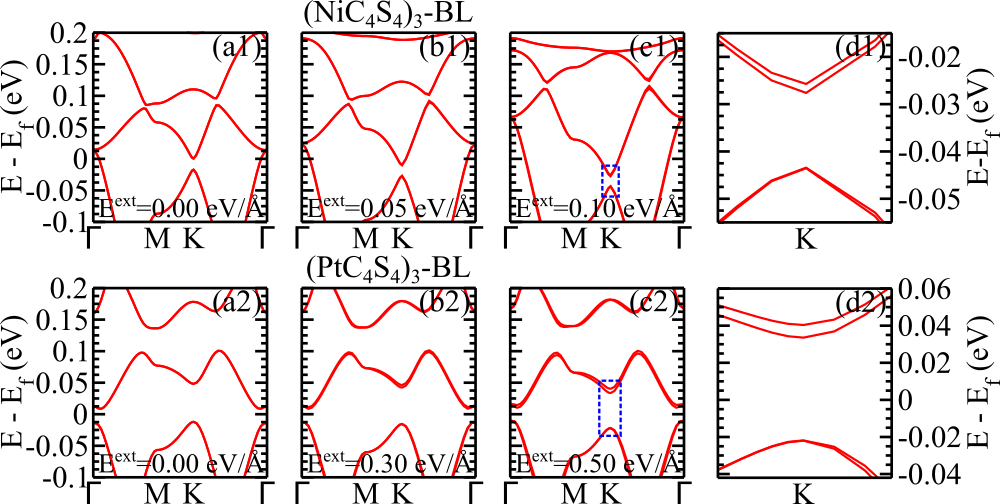}
\caption{\label{bl-fp} First-principles electronic band structure as a function 
of the external electric field E$^{\rm ext}$. (a1)-(c1) {\Ni}-BL and 
(a2)-(c2) {\Pt}-BL. In (d1) and (d2) we evidenciate the blue dashed retangles of (c1) and (c2) respectively.}
\end{figure}

%\bibliography{bib-sup}% Produces the bibliography via BibTeX.
\vspace{10cm}

\hrulefill

[S1] E. Tang, J.-W. Mei, and X.-G. Wen, Phys. Rev. Lett. {\bf 106}, 236802 (2011).
\vspace{0.1cm}

[S2] Z. Wang, Z. Liu, and F. Liu, Phys. Rev. Lett. {\bf 110}, 196801 (2013).
\vspace{0.1cm}

[S3] A. A. Soluyanov and D. Vanderbilt, Phys. Rev. B {\bf 83}, 235401 (2011).
\vspace{0.1cm}

[S4] R. Yu, X. L. Qi, A. Bernevig, Z. Fang, and X. Dai, Phys. Rev. B {\bf 84}, 075119 (2011).
\vspace{0.1cm}

[S5] M. Gmitra, S. Konschuh, C. Ertler, C. Ambrosch-Draxl, and J. Fabian, Phys. Rev. B {\bf 80}, 235431 (2009).
\vspace{0.1cm}

[S6] C. L. Kane and E. J. Mele, Phys. Rev. Lett. {\bf 95}, 146802 (2005).

\end{document}